\documentclass[12pt]{article}

\usepackage{mike,mathptmx,helvet,courier,longtable,graphicx,fullpage}

\setlongtables
\usepackage{apalike}

\newcommand{\ix}[1]{\ensuremath{I_{\mathit{#1}}}}

\newcommand{\jbx}[1]{\ensuremath{\overline{J}_{\mathit{#1}}}}
\newcommand{\pbx}[1]{\ensuremath{\overline{P}_{\mathit{#1}}}}
\newcommand{\px}[1]{\ensuremath{P_{\mathit{#1}}}}
\newcommand{\Cl}{\ensuremath{\mathrm{Cl}^{-}}}
\newcommand{\K}{\ensuremath{\mathrm{K}^{+}}}
\newcommand{\Na}{\ensuremath{\mathrm{Na}^{+}}}

\newlength{\figwidth}
\begin{document}

\markboth{Spike timing precision\ldots}{M. Stiber}

\title{Spike timing precision and neural error correction: local
behavior}

\author{Michael Stiber \\
Computing \& Software Systems \\
University of Washington \\
Bothell, WA, 98011-8246 USA \\
\texttt{stiber@u.washington.edu}}

\maketitle

\begin{abstract}
\normalsize
The effects of spike timing precision and dynamical behavior on error
correction in spiking neurons were investigated.  Stationary
discharges --- phase locked, quasiperiodic, or chaotic --- were
induced in a simulated neuron by presenting pacemaker presynaptic
spike trains across a model of a prototypical inhibitory synapse.
Reduced timing precision was modeled by jittering presynaptic spike
times.  Aftereffects of errors --- in this communication, missed
presynaptic spikes --- were determined by comparing postsynaptic spike
times between simulations identical except for the presence or absence
of errors. Results show that the effects of an error vary greatly
depending on the ongoing dynamical behavior. In the case of phase
lockings, a high degree of presynaptic spike timing precision can
provide significantly faster error recovery. For non-locked behaviors,
isolated missed spikes can have little or no discernible aftereffects
(or even serve to paradoxically reduce uncertainty in postsynaptic
spike timing), regardless of presynaptic imprecision. This suggests
two possible categories of error correction: high-precision locking
with rapid recovery and low-precision non-locked with error immunity.

\end{abstract}

\section{Introduction}

This work is concerned with the effects of spike timing precision and
dynamical behavior on the recovery of ongoing neural discharges after
a perturbation, or an \emph{error}. The concepts of ``precision'' and
``error'' are of course intimately linked. Determining how information
is transmitted among neurons is an essential issue in computational
neuroscience, with a spectrum of possibilities, ranging from overall
average firing rates to timing of individual spikes
\cite{dan-etal98,lestienne01}. Central to the question of where a
presynaptic neuron's code lies in this spectrum is the precision of
spike timing necessary to elicit a particular response from a
postsynaptic neuron. This follows logically, as two different
presynaptic discharges which produce the same postsynaptic response
have achieved the same functional result, and thus can be said to have
carried the same message.

The present paper addresses this issue directly by comparing recovery
from errors in a spike train for two levels of spike timing
precision. The goal is to test the feasibility of the idea that extra
spike timing precision (in the sense that it is more than the minimum
necessary to transmit a given amount of information) can be used for
error correction. The errors that are used are missed spikes, based on
the observed apparent unreliability of synaptic transmission
\cite{katz66,allen-stevens94}.

Examination of the effects of noise and jitter in spike trains on
neural behavior goes back decades
\cite{segundo-etal63,segundo-etal68,bryant-etal73,segundo-etal76,kohn-etal81,kohn-segundo83,segundo-etal94d}.
Briefly, the larger the number of input synapses (the smaller the
effect of each presynaptic spike) and the smaller the correlation
among them, the more their aggregate postsynaptic effect was like a DC
bias. Greater correlations and fewer, stronger synapses operated more
like single, strong synapses, producing reliably repeatable responses
in the postsynaptic cell.  The effect of increasing noise on ongoing
dynamical behavior in pacemakers was observed as steadily narrowing
the input rate domains for phase locked behaviors, abolishing those
with narrower domains before those with broader ones. Consequently,
periodic behaviors with broader domains in parameter space were more
stable in the presence of noise. Later work showed this effect in rat
spinal interneurons \cite{beierholm-etal01}, and presented similar
results for \ratio{1}{1} alternation (the non-pacemaker counterpart
of phase locking) with quasiperiodic driving \cite{tiesinga02}.

Analysis of networks showed that coupling large numbers of cells could
lead to lower output variability in the face of input jitter
\cite{marsalek-etal97,tiesinga-sejnowski01}.  Additionally, periodic,
phase locked dynamics were noted to be optimal for reduction of output
jitter (the exception to this being near bifurcation boundaries, where
input jitter could occasionally abolish locking) in network models
\cite{tiesinga-sejnowski01}.

This paper investigates how timing precision in a presynaptic train
interacts with the destruction of information caused by a single
neuron's internal dynamics. It extends previous work by evaluating
both periodic and aperiodic (quasiperiodic and chaotic) postsynaptic
dynamics, as well as the effects of proximity to bifurcations bounding
periodic behaviors.  Additionally, rather than simply examining the
amount of perturbation produced by presynaptic errors, it also
analyzes the time course of recovery.  The ongoing presynaptic spike
trains used here were stationary ones, with and without jitter, that
induced a range of dynamical behaviors in the postsynaptic cell.
Error recovery was measured by the time course of return of the
postsynaptic cell to its stationary behavior. This was determined by
comparing pairs of simulations that were identical except for the
presence of errors, so that after an error, the spike train of the
postsynaptic cell could be compared to that of one that didn't see the
error. Pairs of simulations were repeated for inputs with and without
jitter, and error recovery was then compared in those two cases.

\section{Methods}

A well-characterized physiological model (see Appendix) of the
crayfish slowly adapting stretch receptor organ, or SAO, was used. In
the living preparation, the SAO produces spike trains with rate
broadly proportionate to muscle stretch; at constant stretch, it
produces pacemaker spike trains.  It includes the recognized prototype
of a moderately-powerful, GABAergic inhibitory synapse, and thus has
been used as a living model to explore synaptic coding across
inhibitory synapses in general.  The model has been used previously to
explain the dynamics behind ``paradoxical accelerations'' observed in
the living preparation \cite{segundo79,buno-etal87} in response to
pacemaker inhibitory PSPs \cite{segundo-etal91a,segundo-etal91b}, as
well as the SAO's responses to nonstationary spike trains
\cite{segundo-etal94a,segundo-etal95c,segundo-etal95d,stiber-etal97b,segundo-etal98}.

Presynaptic spike train timing generation and analysis was performed
in MATLAB~\cite{matlab}. Simulation was performed using custom C code
and the ODEPACK differential equation integration library
\cite{hindmarsh83}. Both pre- and postsynaptic
spike trains were assimilated to point processes~\cite{cox-isham80}
and all analysis was based on inter-event intervals.

In the absence of presynaptic input, the SAO and its model behave as
pacemakers, producing action potentials with an almost invariant
interspike interval, $N$: its \emph{natural} interval. In the presence
of pacemaker presynaptic trains, both exhibit a variety of dynamical
behaviors, including phase lockings, quasiperiodicities, and chaos,
with characteristic dependencies on the presynaptic rate
\cite{segundo-etal91a,segundo-etal91b,stiber-segundo93a,stiber-etal97a}.

The current work began by building on previous works' survey of the
dynamical behaviors exhibited by the model.  Approximately 10,000
simulations were performed, varying both presynaptic rate ($N/I$,
$1/I$ being the presynaptic rate for each simulation) and IPSP
strength ($P$) within biologically plausible limits. This was used to
produce an Arnol'd map, or two-dimensional bifurcation diagram, which
categorized each simulation's dynamical behavior within the 2D $(N/I,
P)$ plane.  Three values of $P$ were then chosen for more detailed
examination. These were dubbed ``weak'', ``moderate'', and ``strong''
IPSPs, based on the ratio of locked (periodic) to non-locked
(aperiodic) behaviors they engendered ($\leq$20\%, $\approx$50\%, and
$\approx$100\%, respectively). For the current investigation, a
moderate value of IPSP strength was chosen.

A range of normalized presynaptic rates $0.2
\leq N/I \leq 2.0$ was then explored using one-dimensional bifurcation
diagrams. Here, such diagrams were constructed by plotting
postsynaptic spike \emph{phase} (cross interval from a postsynaptic
spike back to the most recent presynaptic spike) versus $N/I$. These
allowed easy determination of the distribution and organization of
qualitatively different dynamical behaviors along the presynaptic rate
scale (types of behaviors, behavior ranges, and bifurcation
locations). Each bifurcation diagram involved a set of 90--200
simulations, with diagrams produced for pacemaker and jittered
pacemaker presynaptic spike trains. Values of jitter ($w$, see below)
ranging from $0.01 I$ to $0.1 I$ ($\pm$1\%--$\pm$10\%) were examined;
their bifurcation diagrams were similar overall. Results for $\pm$1\%
jitter are presented here.  This was viewed as a stricter test of the
hypothesis (that high precision inputs would result in faster recovery
from errors than low precision, and thus the extra precision could be
viewed as a type of redundant, error-correcting code) than larger
values of $w$, as larger values would certainly result in the model's
state being perturbed farther from the stationary attractors (and thus
it would be a much shorter distance, therefore faster, for the model
to return to its attractor for high-precision input).  Work on
evaluating, in detail, the effects of varying level of jitter on error
recovery are ongoing \cite{stiber-holderman04b}.

Four sets of 90--100 simulations used to produce bifurcation diagrams
(high precision, low precision, and each with errors, see below) then
served as a database describing error responses of the model. This
communication focuses on the details of the model's recovery after an
error and the effects of presynaptic spike timing precision on that
recovery, and so a few illustrative behaviors were selected. The
behaviors chosen are archetypical of the full range of dynamical
behavior types exhibited by driven nonlinear oscillators:
\ratio{1}{1} locking (period 1), \ratio{1}{2} locking (period 2),
\ratio{2}{3} locking (period 3, the longest periodic behavior for
this model with significant extent along the $N/I$ scale), two
quasiperiodicities (low --- $N/I < 1.0$ --- and high --- $N/I > 1.0$
--- rate), and chaos. An additional behavior was chosen to illustrate
the effects of proximity to a bifurcation.  A more complete
examination of the global behavior of error correction (e.g., two and
three dimensional bifurcation structure) is under way
\cite{stiber-holderman04a}.

\begin{figure}
\centerline{\includegraphics[width=\figwidth]{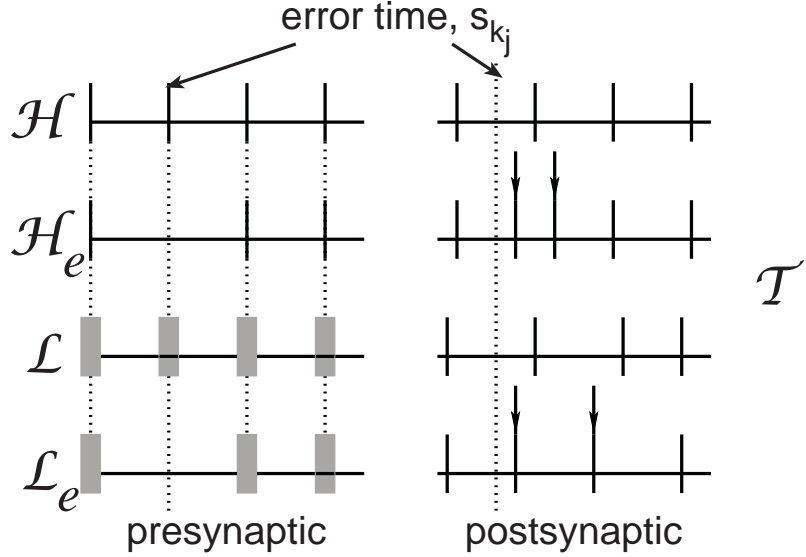}}
\caption{Simulation methods. Pairs of high ($\mathcal{H}$,
$\mathcal{H}_e$) or low ($\mathcal{L}$, $\mathcal{L}_e$) precision
spike trains were applied via the simulated inhibitory fiber
(IF). Postsynaptic spike times ($\mathcal{T}$) were recorded for
analysis.\label{fg:simulation-methods}}
\end{figure}

For each example behavior, simulations involved generating two pairs
of almost identical presynaptic spike trains, as schematized in
Figure~\ref{fg:simulation-methods}. For each set of simulations, a
reference, high precision train, $\mathcal{H}$, with spike times
$\{s_1, s_2, \ldots, s_n\}$, was first generated. As this was a
pacemaker train, all interspike intervals were identical ($\forall k,
I = I_k = s_k - s_{k-1}$). $I$ was the independent variable, chosen to
elicit the example dynamical behaviors from the model. An \emph{error}
was defined to be complete failure of synaptic transmission, and so a
set of spike numbers, $\mathcal{K}= \{k_1, k_2,
\ldots, k_m\}$, was generated randomly such that all errors were
separated by at least 15s (to allow the neuron to return to its
stationary behavior before each error). Presynaptic spikes at times
$\mathcal{H}_\mathcal{K} = \{s_{k_1}, s_{k_2}, \ldots, s_{k_m}\}$ were
eliminated from $\mathcal{H}$ to produce a high precision
\emph{erroneous} train, $\mathcal{H}_e=\mathcal{H} -
\mathcal{H}_\mathcal{K}$.

A low precision counterpart of $\mathcal{H}$ was generated by
jittering each spike $s_k \in \mathcal{H}$ by a value $u_k$ taken from
a uniform distribution with range $\pm wI$. In the current
communication, a small value of $w$ ($w=0.01$, 2\% jitter) was
used.

The result of applying this jitter was a low precision train
$\mathcal{L}=\{s_1+u_1, s_2+u_2, \ldots, s_n+u_n\}$. A low precision
erroneous train, $\mathcal{L}_e=\mathcal{L} -
\mathcal{L}_\mathcal{K}$, was then produced by eliminating the spikes
numbered $\mathcal{K}$ at times $\mathcal{L}_\mathcal{K} =
\{s_{k_1}+u_{k_1}, s_{k_2}+u_{k_2}, \ldots,
s_{k_m}+u_{k_m}\}$. Subsequent analysis involved pooling timing
data for postsynaptic spikes near errors in these sets of four
simulations. As a double-check, multiple simulations were run for each
set of simulation parameters (with different realizations of
$\{u_k\}$) and analysis performed on the pooled data from all; as
might be expected, there was no difference in results for $m$ errors
taken from $n$ simulations versus $m$ errors taken from a single
simulation.

\begin{figure}
\centerline{\includegraphics[width=\figwidth]{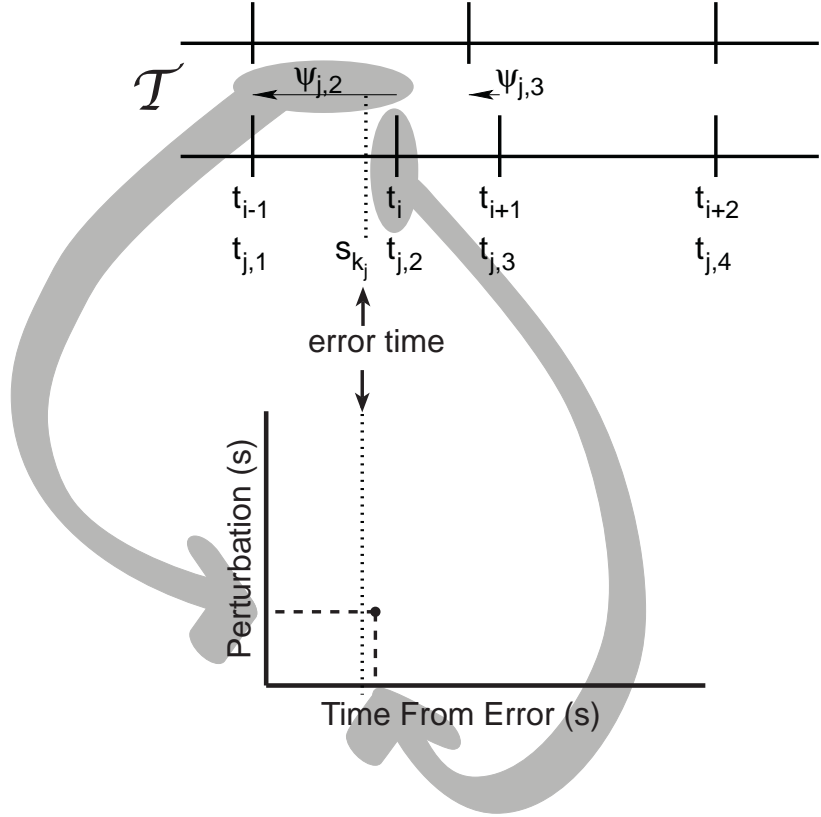}}
\caption{Perturbation graph generation. The displacement of a
postsynaptic spike after an error (bottom train) from the time of the
corresponding spike in the situation where there was no error (top
train) is plotted versus time since the error.\label{fg:perturb-methods}}
\end{figure}

As Figure~\ref{fg:perturb-methods} illustrates, the first step of
analysis was based on comparing times of postsynaptic spikes
$\mathcal{T} = \{\ldots, t_{i-1}, t_i, \ldots\}$ just before and after
the error times for the erroneous and error-free cases. For each error
$k_j \in \mathcal{K}$, $j = 1, 2, \ldots, m$, the set of $c$
postsynaptic spikes just before and after were selected (here, 5
spikes before and 25, 30, or 45 spikes after --- depending on the rate
of production of postsynaptic spikes and the recovery time --- and
thus $c$ was 30, 35, or 50). These spike times can be renamed relative
to the error number $j$ as $\{t_{j,1}, t_{j,2},
\ldots, t_{j,c}\}$ (bottom train in figure). For each postsynaptic
spike in the erroneous case, $\{t_{j,g}\}$, $j = 1, 2, \ldots, m$, $g
= 1, 2, \ldots, c$, the cross interval back to the most recent
postsynaptic spike in the error-free case (top train in figure),
$\psi_{j,g}$, was computed. As the simulations involve an inhibitory
synapse, the number of postsynaptic spikes in the erroneous case was
always greater than or equal to the number in the error-free case;
this was the reason for using the former spikes as the reference for
computing $\psi_{j,g}$.  This was done for each pair of high and low
precision input simulations to yield $\psi^\mathcal{H}_{j,g}$ and
$\psi^\mathcal{L}_{j,g}$, respectively. Thus, $\psi^\mathcal{H}_{j,g}$
and $\psi^\mathcal{L}_{j,g}$ measure the shift, or
\emph{perturbation}, of each postsynaptic spike from the time it would
have occurred if there had been no error. The graph of
$(t_{j,g}-s_{k_j},
\psi^\mathcal{H}_{j,g})$ or $(t_{j,g}-s_{k_j}-u_{k_j},
\psi^\mathcal{L}_{j,g})$, such as in Figure~\ref{fg:2to3}(A), was termed
a \emph{perturbation graph}.

In every case, the perturbation graph compares a simulation (either
high or low precision) with errors to the identical (canonical) one
without errors. Therefore, any scatter of points is due to either the
simulation accuracy (which, for lockings at least, can be judged by
the points before the error time) or the presence of errors themselves.
Jitter, in and of itself, did not produce any scatter because each
presynaptic spike in both simulations occurred at identical times in
identical simulations (neglecting errors). Perturbation graphs show
only the effects of errors and possible interaction of errors with
the presence of jitter.

\begin{figure}
\centerline{\includegraphics[width=\figwidth]{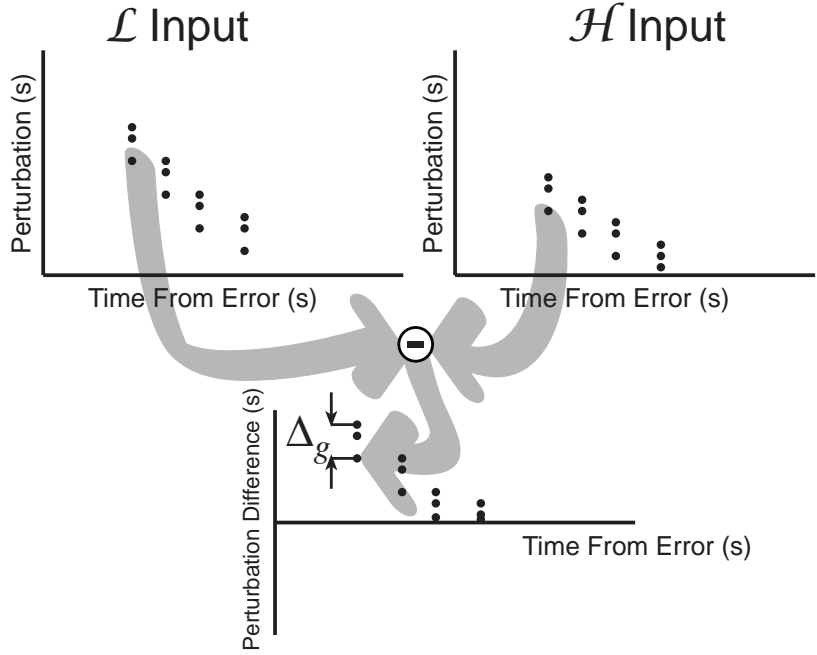}}
\caption{Recovery plot construction. A pair of corresponding high and
low precision perturbation graphs (top) are subtracted to produce
perturbation differences; the recovery plot is the plot of the range
of differences, $\Delta_g$.\label{fg:recovery-methods}}
\end{figure}

The high and low precision perturbations, $\psi^\mathcal{H}_{j,g}$ and
$\psi^\mathcal{L}_{j,g}$, were compared in \emph{recovery
plots}. Figure~\ref{fg:recovery-methods} illustrates how such plots
were generated. For each pair of cross intervals, the difference,
$\delta_{j,g} = \psi^\mathcal{L}_{j,g} - \psi^\mathcal{H}_{j,g}$, was
computed. Because of the variation in the individual error responses
in the low precision case (the scatter of points in the perturbation
plot), $\delta_{j,g}$ typically took on a range of values for each
$g$. Rather than examine the average difference, the decision was made
to compute the range of differences, $\Delta_g = \max_j(\delta_{j,g})
- \min_j(\delta_{j,g})$. A difference $\delta_{j,g} = 0$ indicated
that the high and low precision perturbations, \emph{for that
particular error $j$ and postsynaptic spike $g$}, were identical. A
range $\Delta_g = 0$ indicated that, for \emph{all} errors, the
perturbation of postsynaptic spike $g$ was the same for high and low
precision --- that low precision input did not have an effect on the
perturbation.  Thus, a range $\Delta_g > 0$ indicated that some low
precision perturbations differed from high precision ones. In other
words, because these were simulations, none of the postsynaptic spikes
were considered outliers.  As an alternative, the variance of
$\delta_{j,g}$ (for each value of $g$) could have been used; in
practice, it produced equivalent results.  So, a recovery plot such as
in Figure~\ref{fg:1to1} is a graph of $\Delta_g$ versus the mean
offset $e_g = \overline{t_{j,g}-s_{k_j}}$, for each value of $g = 1,
2,
\ldots, c$. It collapses the
\spscript{$g$}{th} set of perturbations to a single point and compare
high and low precision perturbations, showing the time course by which
the worst low precision error responses approached those for high
precision.

The primary motivations behind the use of recovery plots were easier
analysis of the time course of recovery than perturbation graphs and
elimination of the need to compare pairs of perturbation plots by eye
to evaluate the effects of low versus high precision.  Assuming the
erroneous simulation fully recovered before each error, the values of
$\Delta_g$ for $e_g<0$ can also serve to test whether low precision
responses were still significantly perturbed. Comparing $\Delta_g$ for
$e_g<0$ for different simulation parameters allows one to judge
whether the simulation was recovered.

To characterize the ongoing dynamical behavior exhibited by the SAO,
cross intervals between the pre- and postsynaptic times, $\mathcal{H}$
and $\mathcal{T}$, respectively, were used. Since $\mathcal{H}$ was a
pacemaker process, the cross intervals $\phi_i = t_i - s_k$ from
postsynaptic spikes back to most recent preceding presynaptic spikes
were termed \emph{phases}~\cite{winfree80,glass-mackey88}. The
\emph{phase return map}, plotting $(\phi_i, \phi_{i+q})$, trivially
indicates
\ratio{p}{q} locking since $\phi_i = \phi_{i+q}$. Quasiperiodic
responses produce a 1D curve without a maximum for $q=1$, while
chaotic behaviors have return maps with maxima or that are not
one-dimensional curves~\cite{abraham-shaw84,rapp-etal85,ruelle89}. For
brevity's sake, return maps are omitted from the \emph{Results}
section for phase lockings.

\section{Results}

In this section, error recovery during locked, quasiperiodic, and
chaotic responses is considered. A final subsection examines the
effect of proximity to a bifurcation boundary on error recovery.

\subsection{Phase Lockings}

\begin{figure}
\begin{center}
\begin{tabular}{rl}
\raisebox{1.75in}{\textsf{\large A}} &
\includegraphics[width=2.8in]{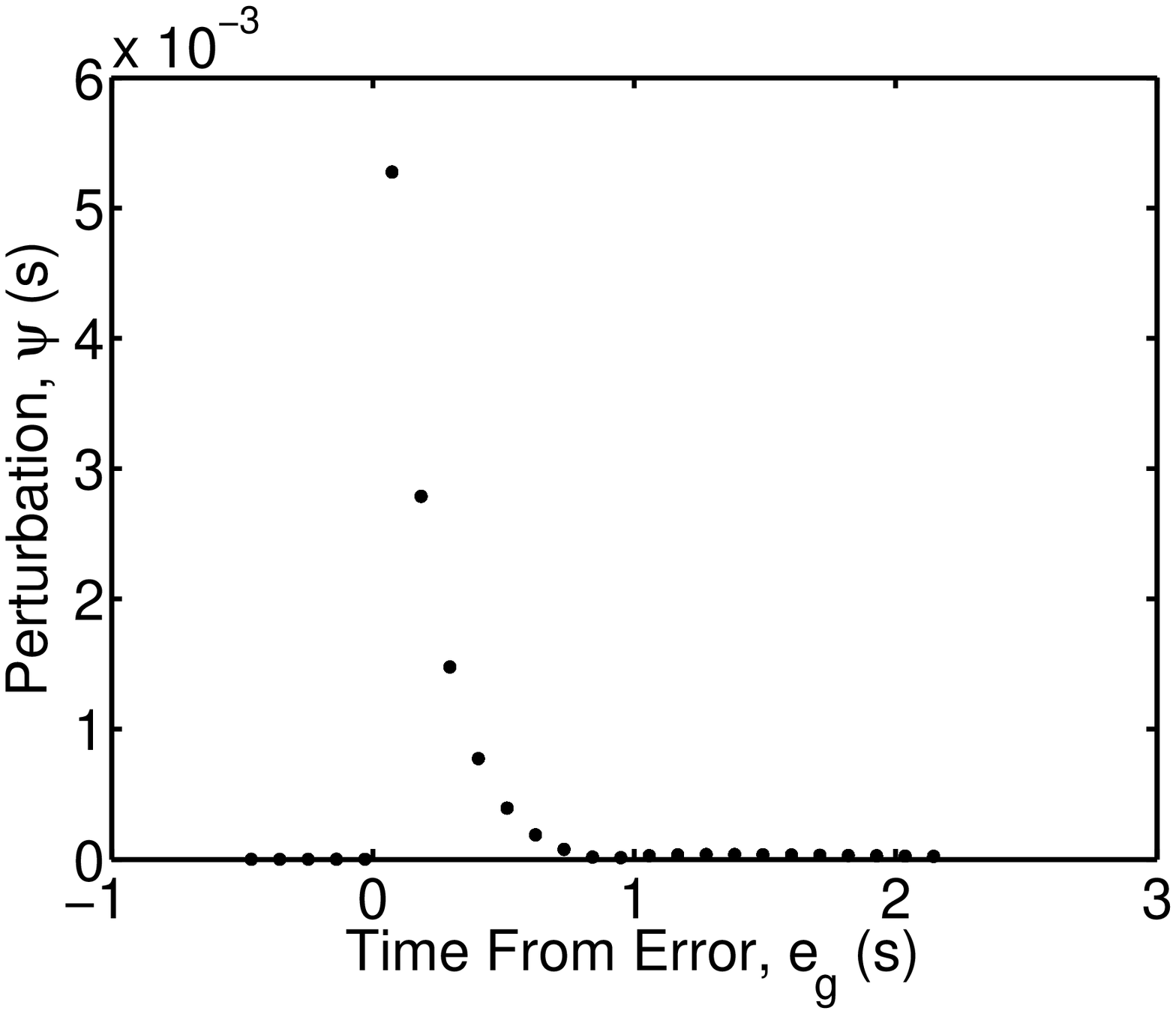} \\
\raisebox{1.75in}{\textsf{\large B}} &
\includegraphics[width=2.8in]{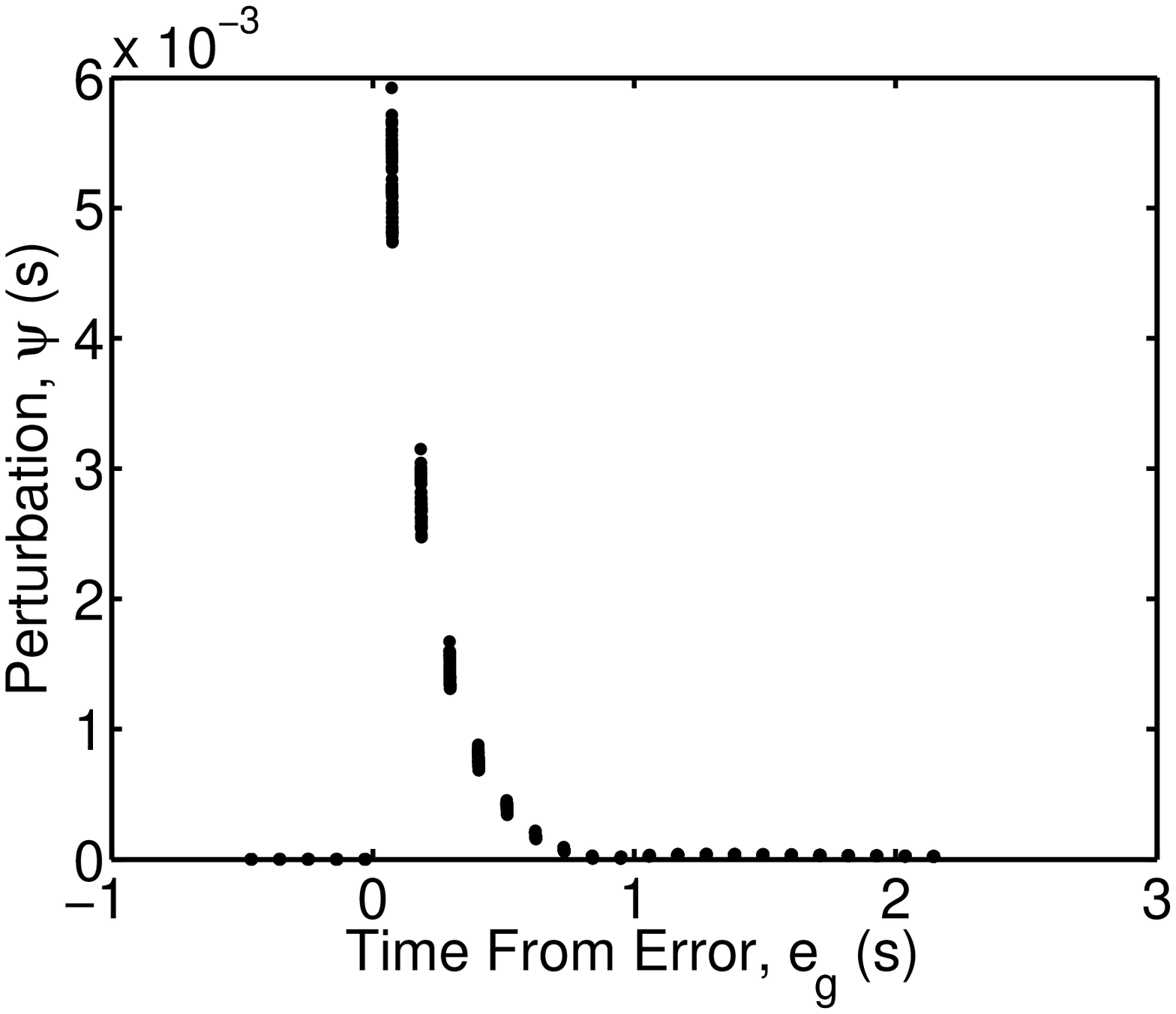} \\
\raisebox{1.75in}{\textsf{\large C}} &
\includegraphics[width=2.8in]{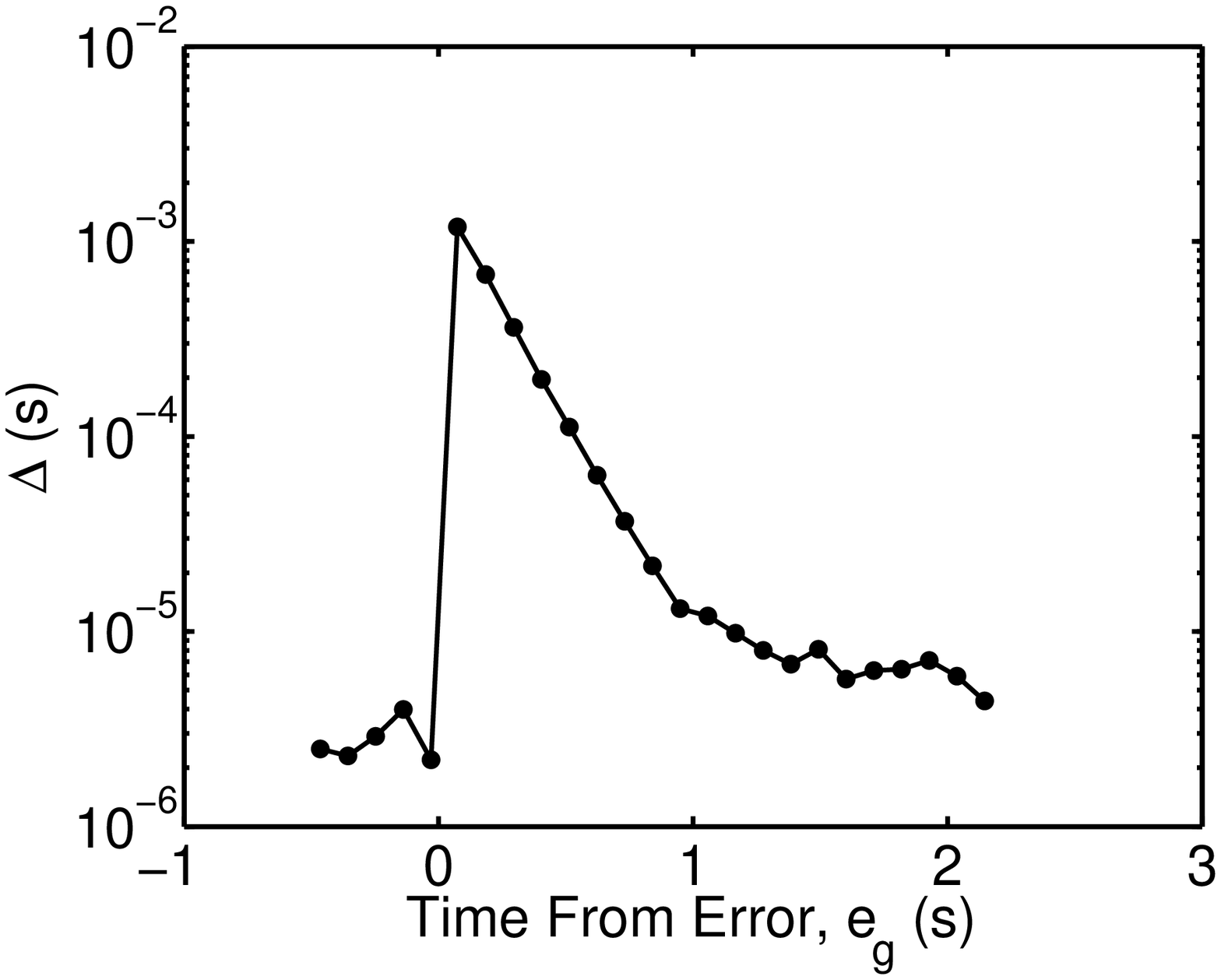}
\end{tabular}
\end{center}
\caption{Results for center of \ratio{1}{1}
locking ($I=108.78$ms, $N/I=0.96$). High precision perturbation graph
(A) shows identical responses to all errors; low precision (B) shows
range of responses. Recovery plot (C) shows long-duration
recovery. ($m=49$ errors, $c=25$ spikes/error)\label{fg:1to1}}
\end{figure}

In previous work on both the SAO and this model, phase locked
responses were found to be the most robust in the presence of
nonstationary presynaptic spike trains, with \ratio{1}{1} locking
being the most robust among
lockings~\cite{segundo-etal93a,segundo-etal94a,segundo-etal95c,segundo-etal95d,segundo-etal98,stiber-etal97b}.
This has also been noted in other preparations and simulations
\cite{beierholm-etal01,tiesinga-sejnowski01,tiesinga02}.  Thus,
analysis began with these behaviors in the expectation that they would
exhibit the greatest immunity to low precision jitter and the fastest
error recovery times~\cite{stiber03a}.

\begin{figure}
\begin{center}
\begin{tabular}{rl}
\raisebox{1.75in}{\textsf{\large A}} &
\includegraphics[width=2.8in]{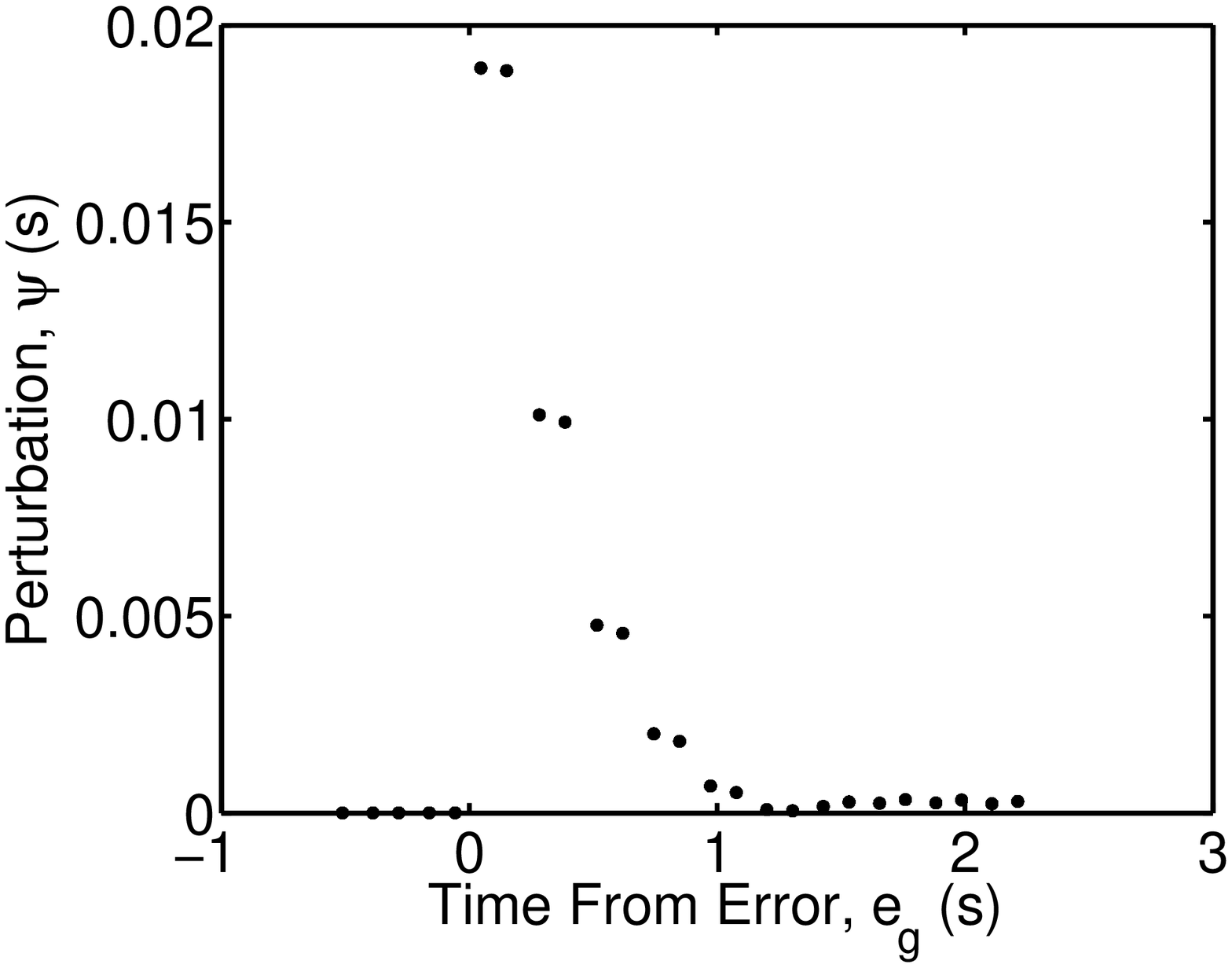} \\
\raisebox{1.75in}{\textsf{\large B}} &
\includegraphics[width=2.8in]{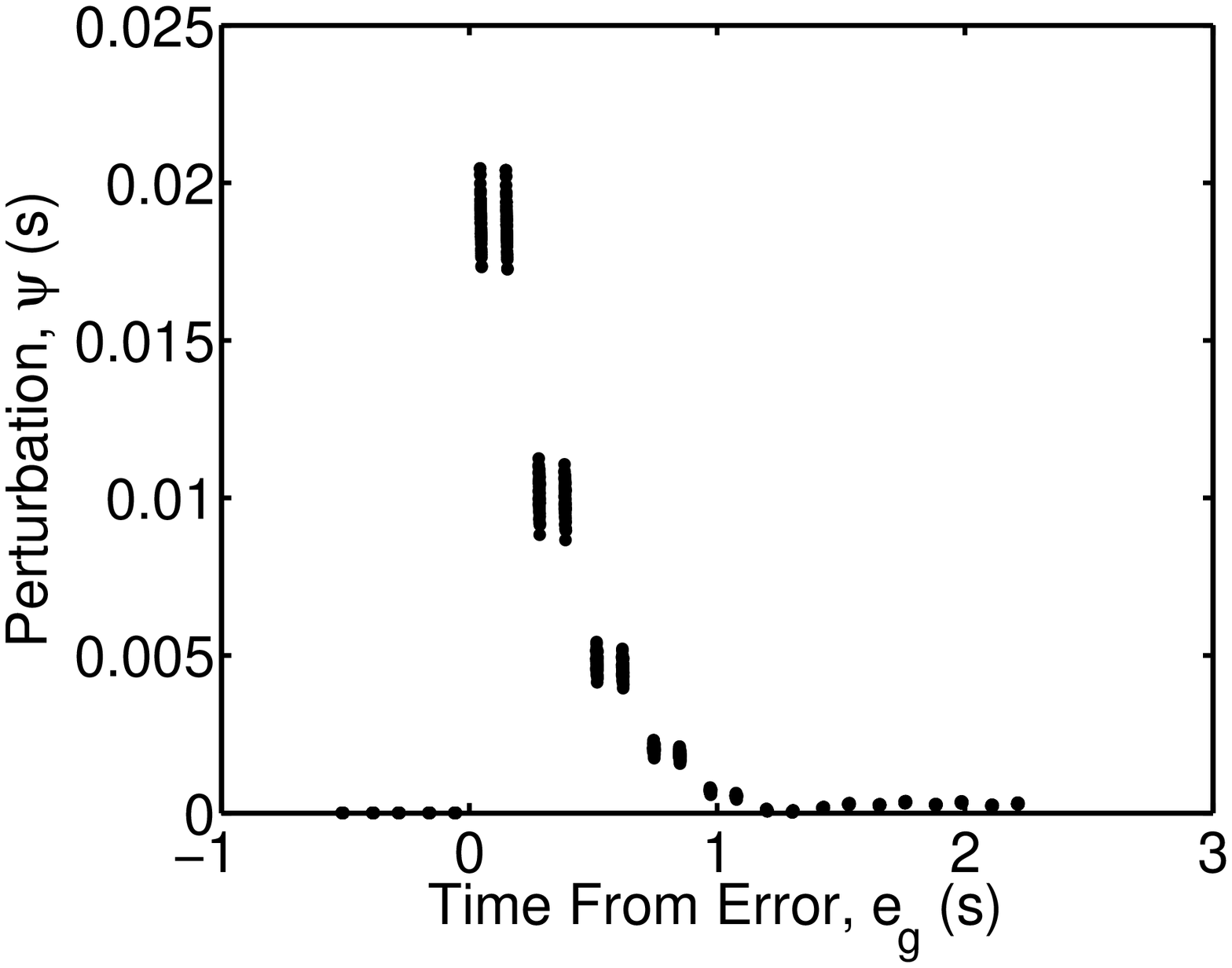} \\
\raisebox{1.75in}{\textsf{\large C}} &
\includegraphics[width=2.8in]{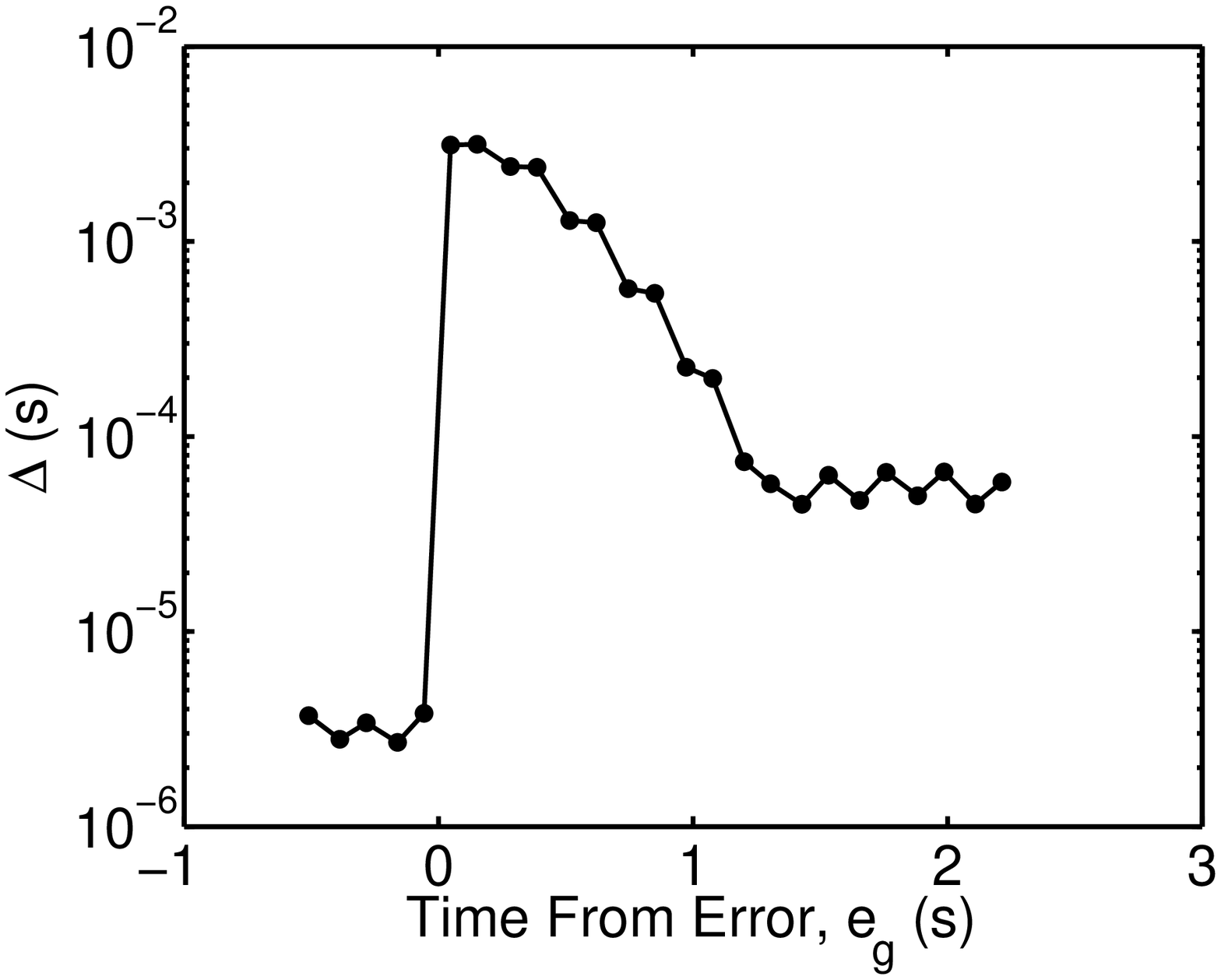}
\end{tabular}
\end{center}
\caption{Results for \ratio{1}{2}
locking ($I=227.01$ms, $N/I=0.46$). Perturbation graphs for high (A)
and low (B) precision input and recovery plot (C) all carry mark of
the two categories of phases. ($m=49$errors, $c=25$
spikes/error)\label{fg:1to2}}
\end{figure}

Figure~\ref{fg:1to1} presents typical results for
\ratio{1}{1} locking. For high precision input, the response to every
error was identical (A), with a maximum perturbation of approximately
$\psi^\mathcal{H}_{j,g}=5.25$ms. For low precision (B), the maximum
perturbation was $\psi^\mathcal{L}_{j,g}
\approx 6$ms, with up to a 1.2ms range for any single value of $g$. An
exponential fit for the points in the recovery plot (C) in the range $0.4
\leq e_g \leq 0.9$ yielded a recovery time constant of $\tau=0.20$s
($\tau/I=1.84$), with a slower recovery proceeding thereafter.  This
is the rate at which the ``worst'' low precision error responses
approached that for high precision.  Low precision responses were
still significantly perturbed 2s after an error.

Perturbation graphs (A, B) and a recovery plot (C) for \ratio{1}{2}
locking are shown in Figure~\ref{fg:1to2}. Note that the points occur
in pairs, with small differences between the pairs of points leading
to the appearance of oscillations in the recovery plot. These
correspond to the pairs of postsynaptic spikes that occur between each
pair of presynaptic spikes (two categories of cross intervals). As
with
\ratio{1}{1} locking, all responses to errors for high precision input
were identical (A); the maximum perturbation in this case being
19ms. Low precision (B) maximum perturbations ranged from about 17ms
to 20ms. The recovery time constant for $0.51 \leq e_g \leq 1.08$ was
$\tau=0.27$s ($\tau/I=1.19$). Recovery slows down considerably after
$e_g=1.2$s, at about ten times the pre-error difference range.

\begin{figure}
\begin{center}
\begin{tabular}{rl}
\raisebox{2in}{\textsf{\large A}} &
\includegraphics[width=2.8in]{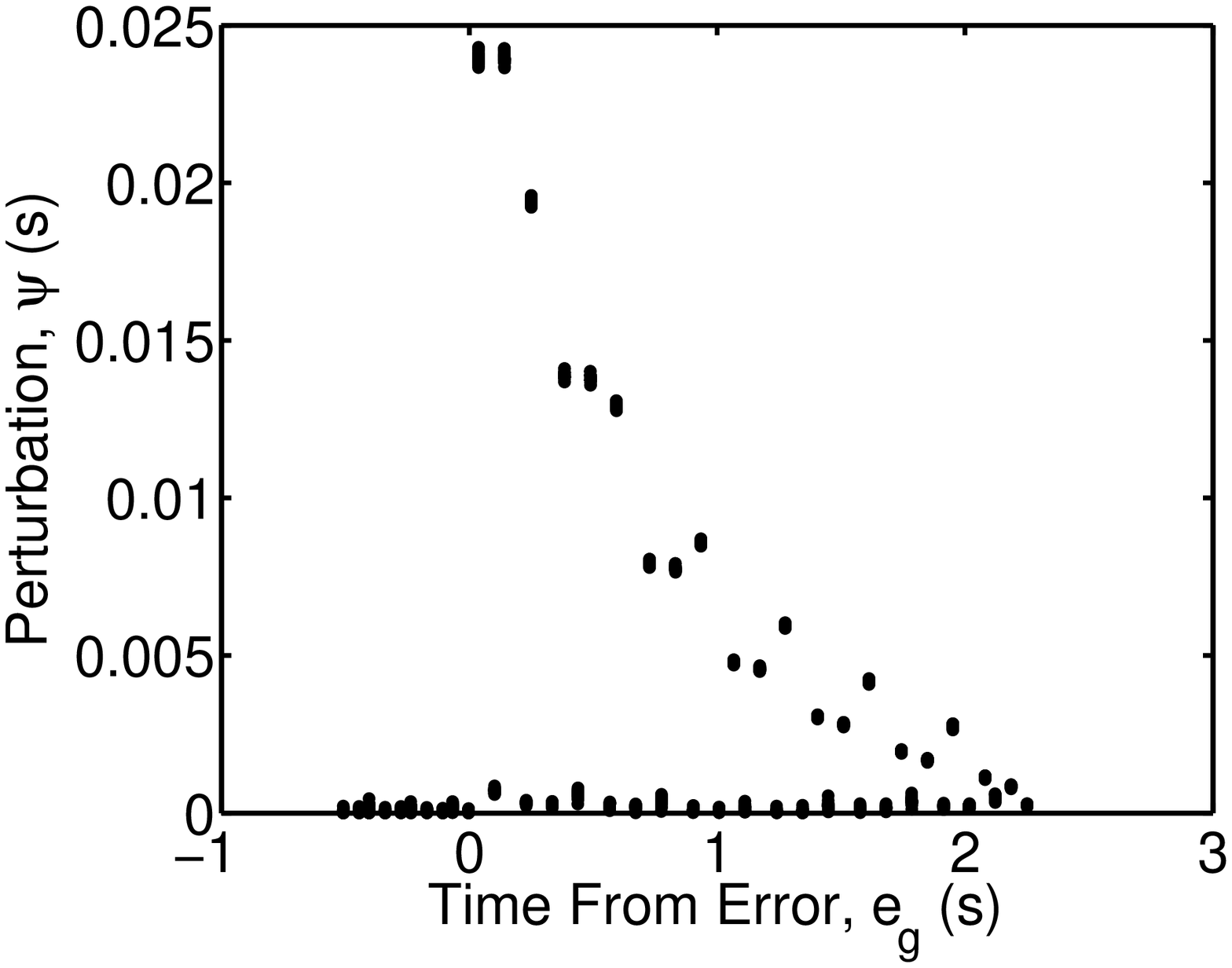} \\
\raisebox{2in}{\textsf{\large B}} &
\includegraphics[width=2.8in]{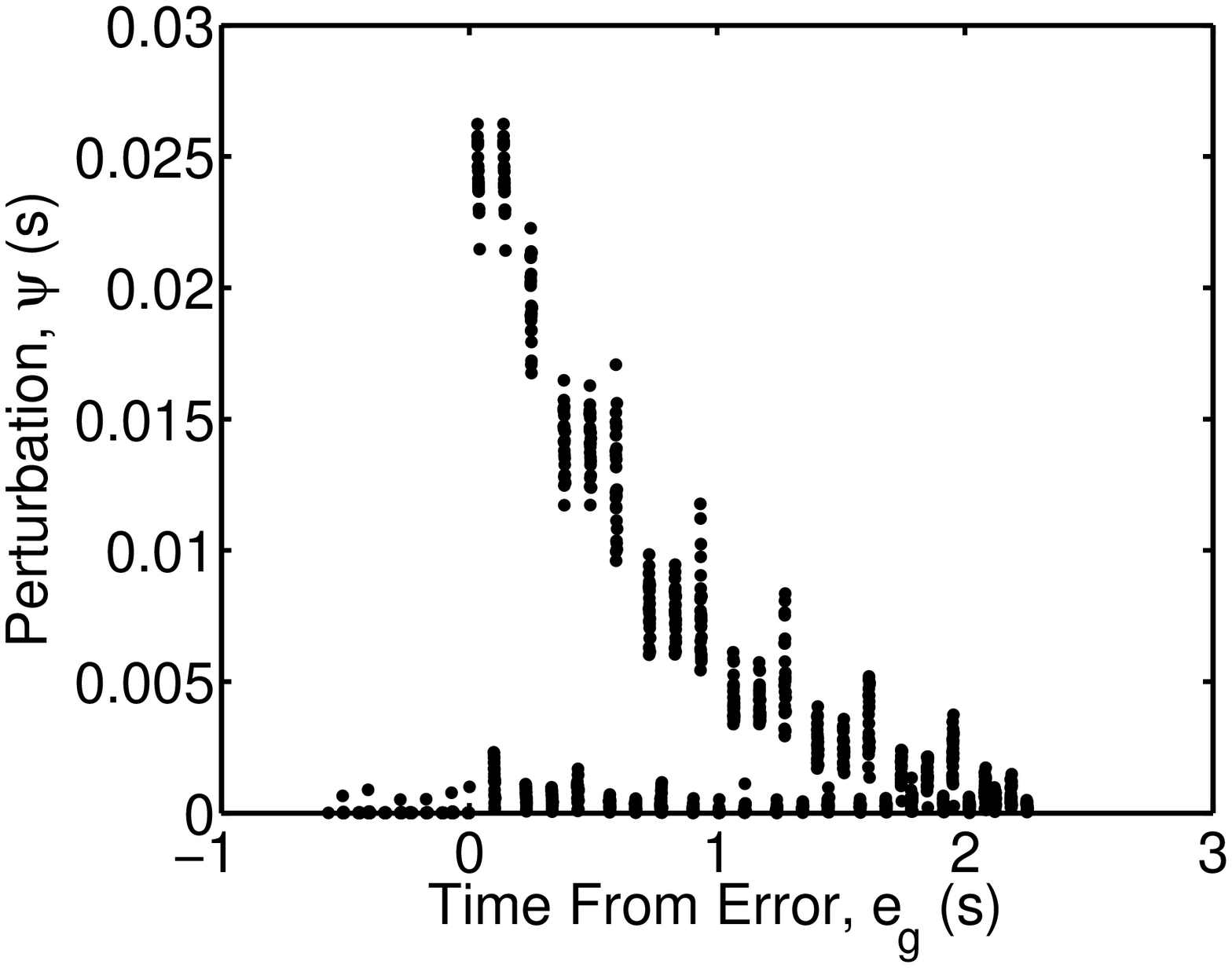} \\
\raisebox{2in}{\textsf{\large C}} &
\includegraphics[width=2.8in]{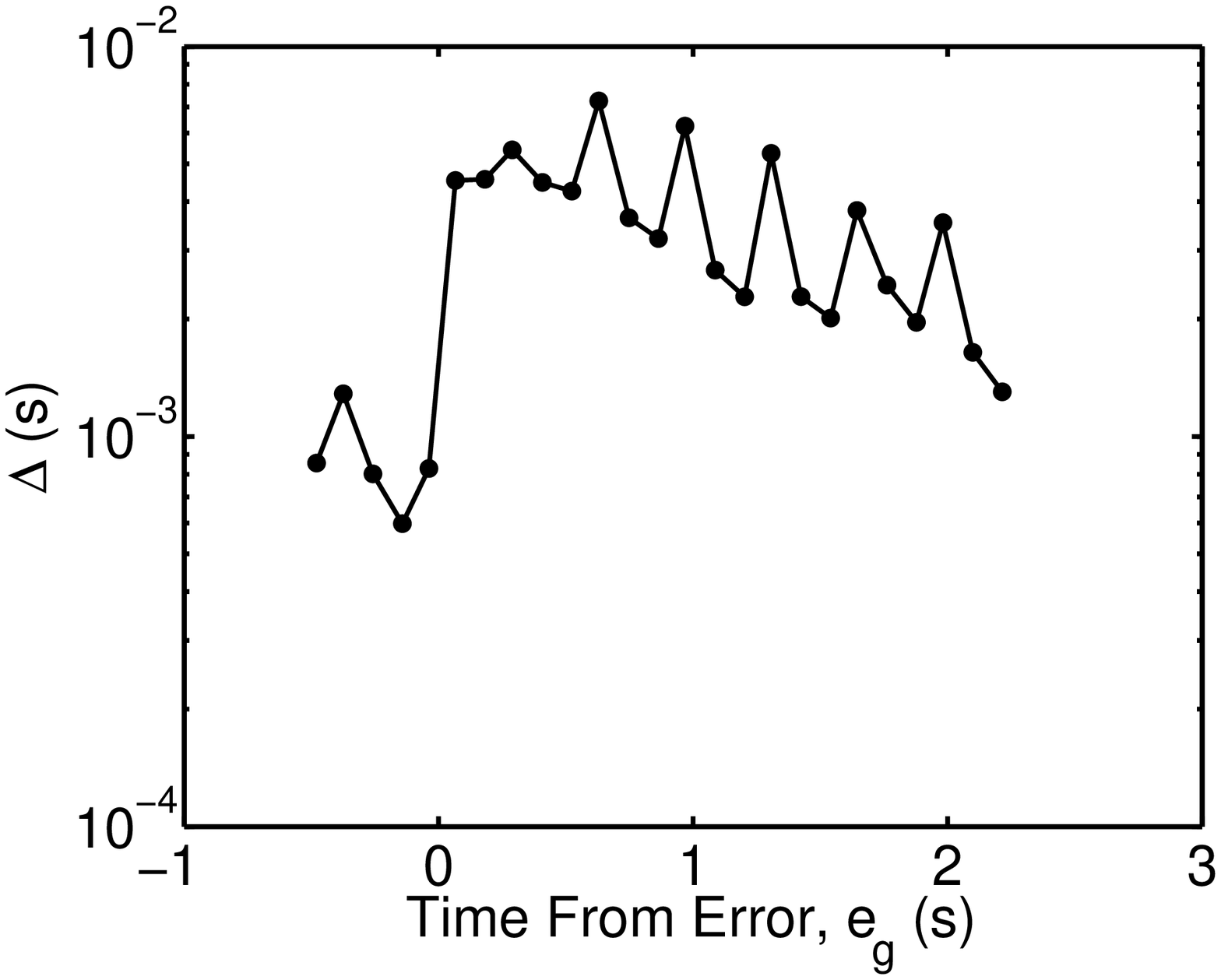}
\end{tabular}
\end{center}
\caption{Results for \ratio{2}{3} locking ($I=168.43$ms,
$N/I=0.62$). Both high precision (A) and low precision (B)
perturbation graphs show ranges of responses to different
errors. Recovery plot (C) shows reduced, long-duration error
aftereffect. ($m=49$ errors, $c=25$ spikes/error)\label{fg:2to3}}
\end{figure}

As Figure~\ref{fg:2to3}(A) shows, \ratio{2}{3} locking, unlike the
previous lockings, exhibited some variability even in its high
precision response to repeated errors --- qualitatively similar to low
precision simulations, though much smaller. Note the linear scale in
graphs (A) and (B).  The maximum high precision error perturbation was
around 24ms with a 1ms spread; for low precision, this was 27ms with a
5ms spread (ignoring bimodality; see below). In both cases, there were
triplets of clusters, corresponding to the three postsynaptic spikes
occurring for each pair of presynaptic spikes (i.e., this was a
period-three behavior). Differences among the three categories of
cross intervals explain what appear to be oscillations in the recovery
plot (C), as with \ratio{1}{2} locking. Clusters had a bimodal
distribution of perturbations, with some small (around 0--2ms) and
some large. The recovery plot shows an initial peak at less than an
order of magnitude more than the pre-error value, a slow recovery
until around $e_g=2.5$s; there were additional, lower peaks after 2.5s
(before eventual return to pre-error values, as indicated by the
points before the error). The range of differences before the error
($e_g<0$) was on the order of 1ms, compared to 2--5$\mu$s for
\ratio{1}{1} and \ratio{1}{2} lockings.

\subsection{Quasiperiodicities}

\begin{figure}
\begin{center}
\begin{tabular}{lclc}
\raisebox{1.35in}{\textsf{\large A}} &
\includegraphics[width=2in]{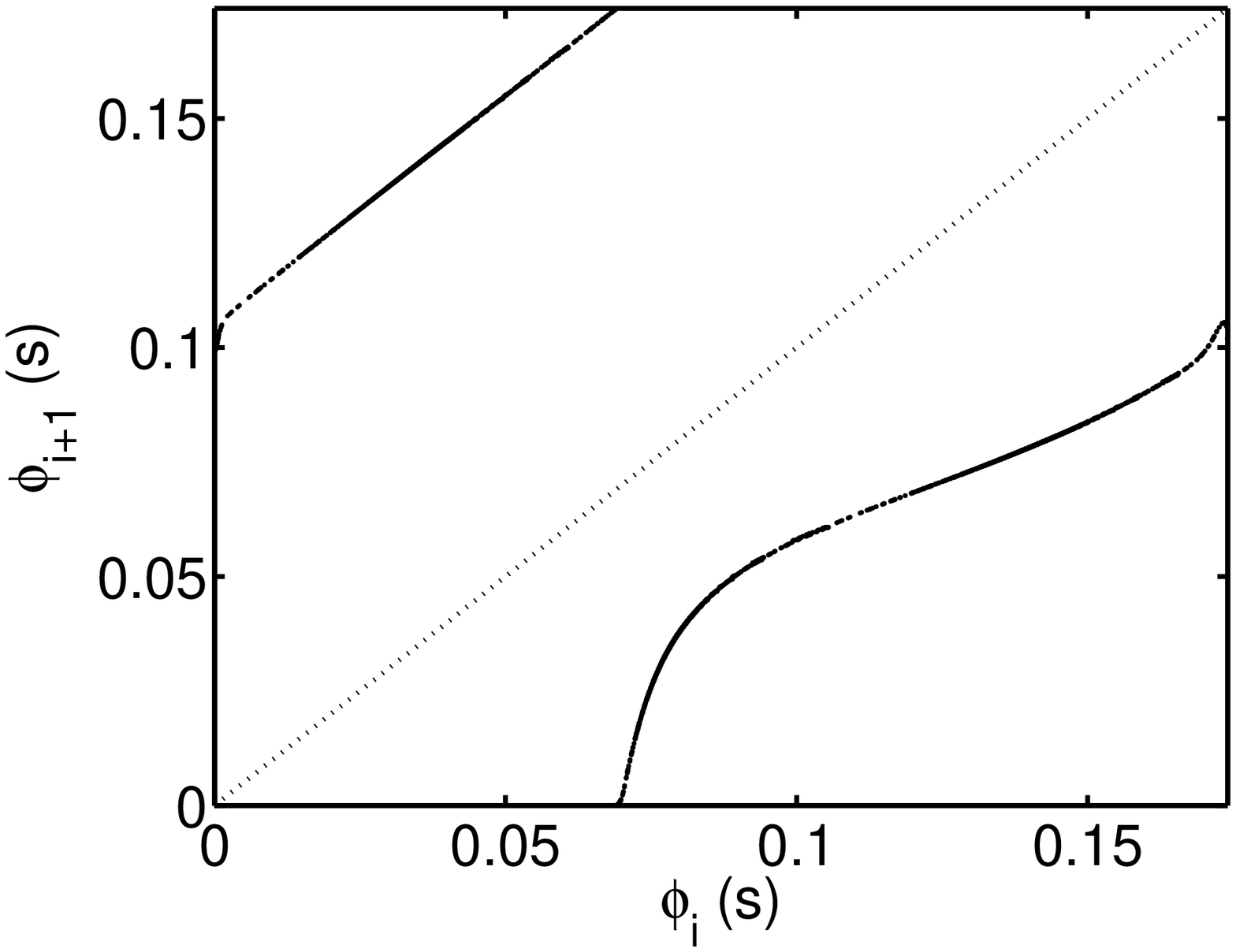} &
\raisebox{1.35in}{\textsf{\large B}} &
\includegraphics[width=2in]{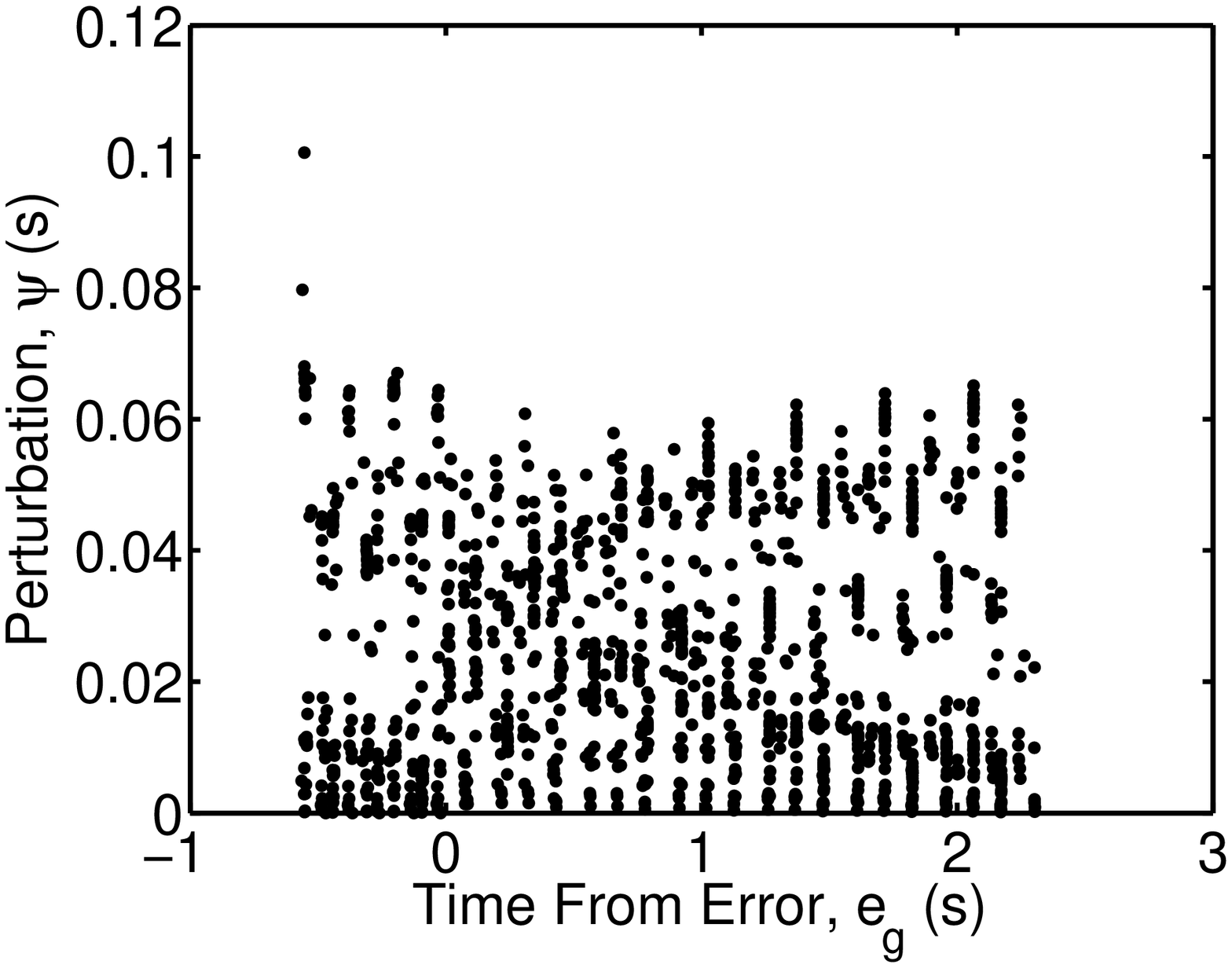} \\
\raisebox{1.35in}{\textsf{\large C}} &
\includegraphics[width=2in]{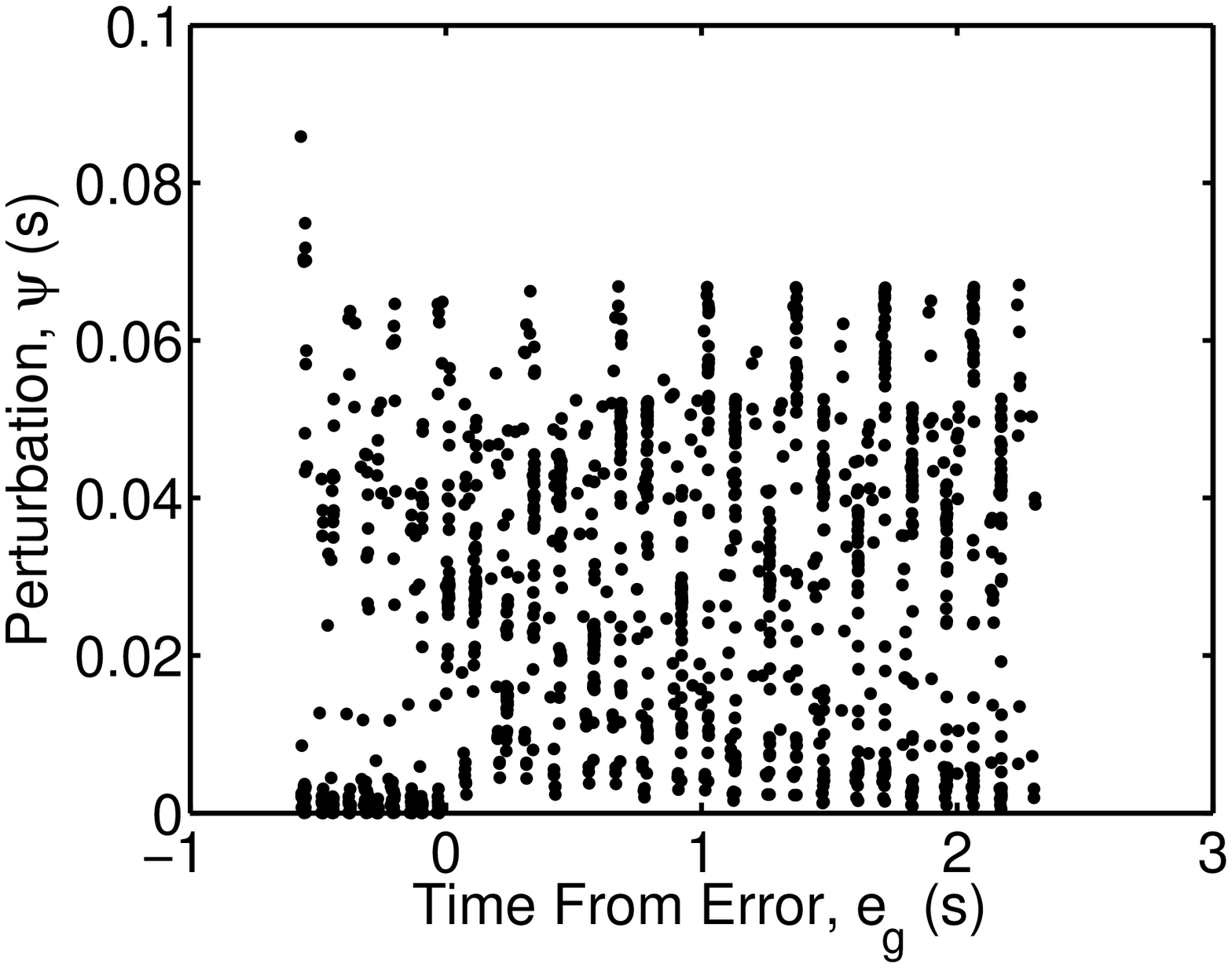} &
\raisebox{1.35in}{\textsf{\large D}} &
\includegraphics[width=2in]{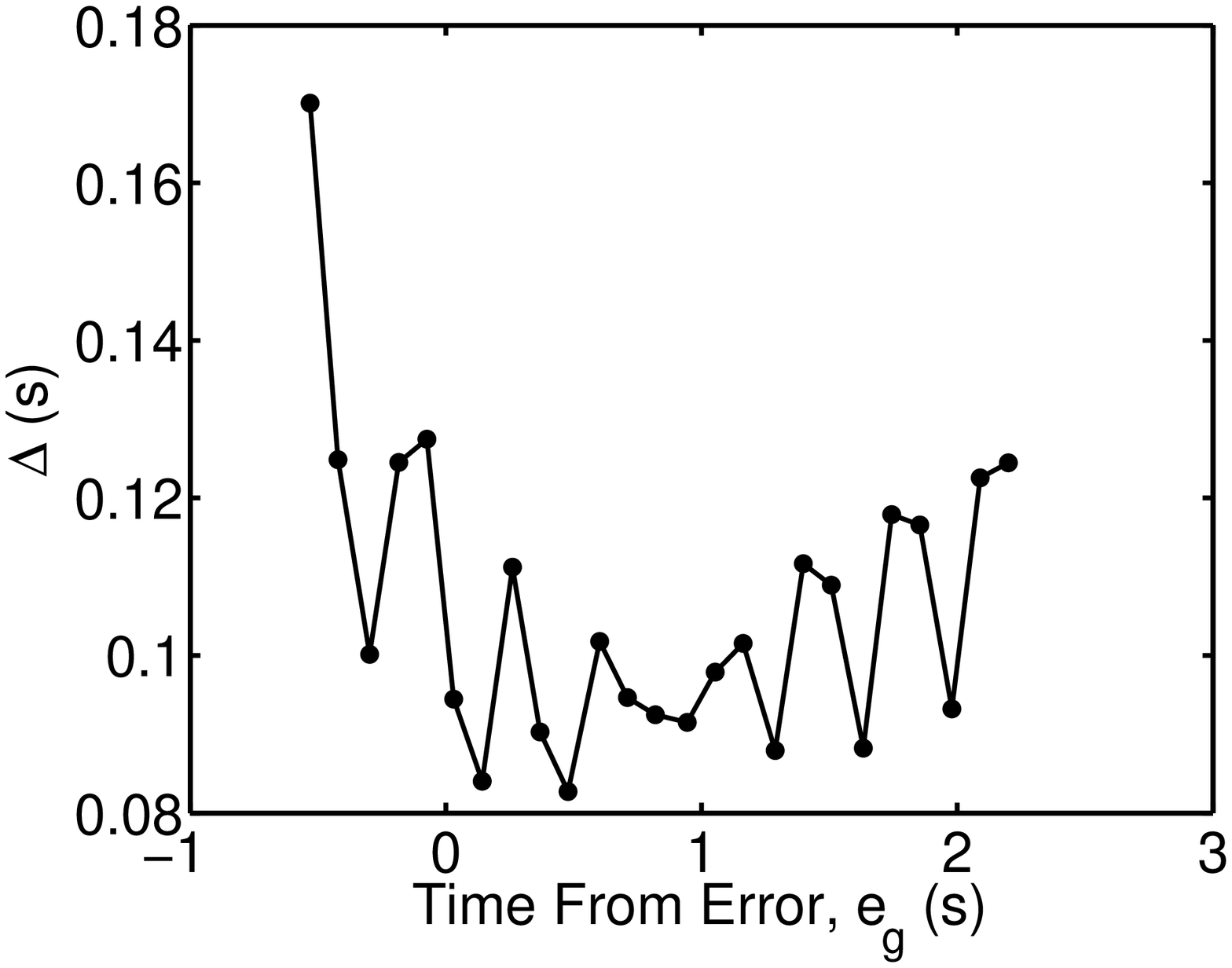}
\end{tabular}
\end{center}
\caption{Error responses for quasiperiodic behavior near \ratio{2}{3}
locking ($I=174.04$ms, $N/I=0.60$). Phase return map (A) is a 1D
curve. High precision (B) and low precision (C) perturbation
plots. Note linear Y axis for recovery plot
(D). ($m=49$ errors, $c=25$ spikes/error)\label{fg:quasiperiodic}}
\end{figure}

A quasiperiodic behavior is one in which the phase of the response
shifts with respect to the stimulus, eventually taking on any value
within some range~\cite{ermentrout-rinzel84}. This is made apparent by
the phase return map in Figure~\ref{fg:quasiperiodic}(A), which is a
one-dimensional curve covering the range 0--$I$. Note that a phase of
zero is equivalent to one of $I$, so the graph corresponds to a torus
and the curve is not discontinuous. There was no apparent qualitative
difference (and little quantitative difference) between high (B) and
low (C) precision error responses; if anything, there was even a
slight decrease in postsynaptic spike timing variability after an
error. This is especially apparent in the recovery plot (D), with a
maximum range of $I$ before an error and a minimum range (ignoring the
high frequency ``oscillations'' every third spike --- the ``ghost'' of
the nearby \ratio{2}{3} locking) of around 95ms at $e_g \approx 1$s.

\begin{figure}
\begin{center}
\begin{tabular}{lclc}
\raisebox{1.35in}{\textsf{\large A}} &
\includegraphics[width=2in]{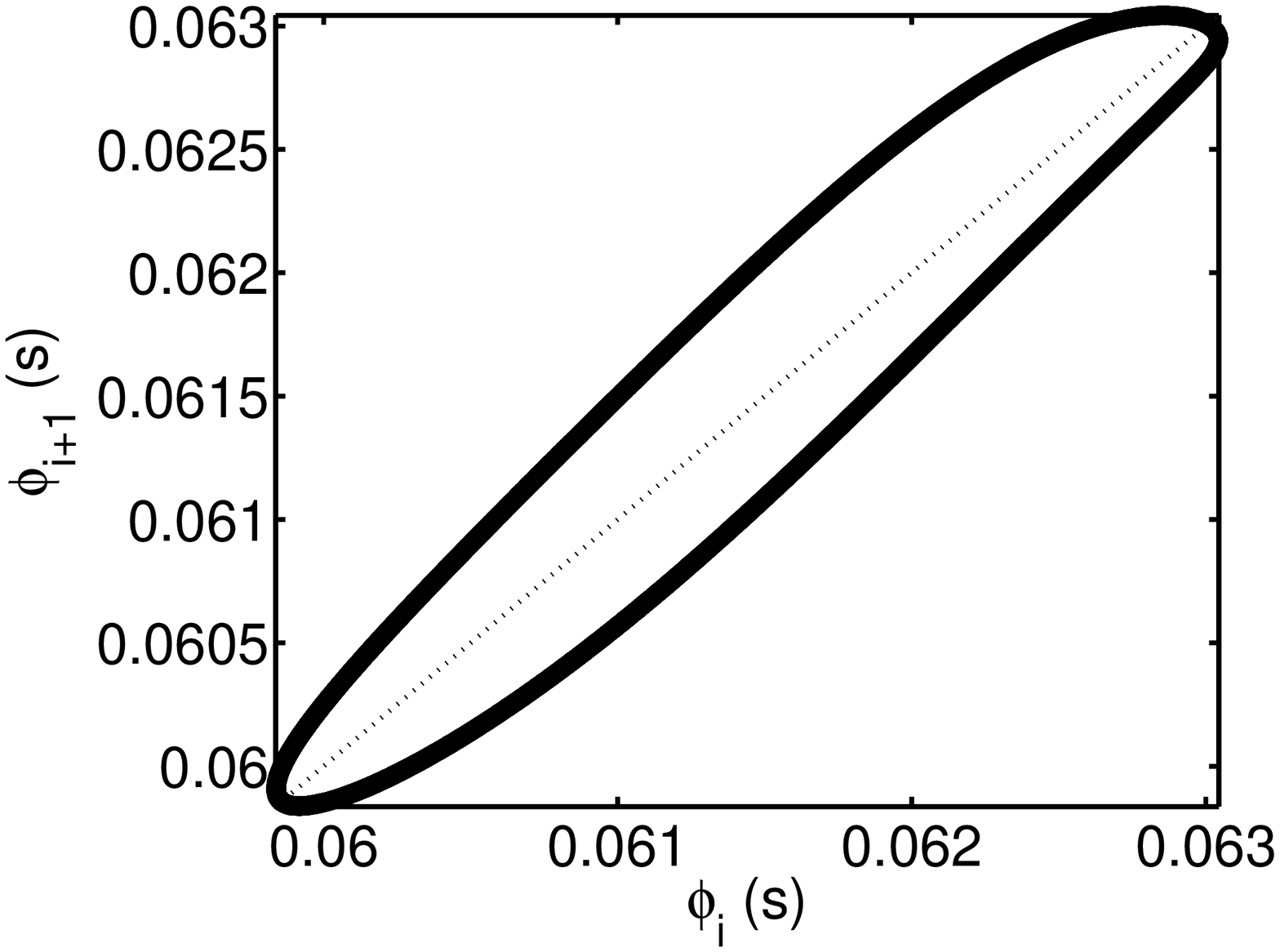} &
\raisebox{1.35in}{\textsf{\large B}} &
\includegraphics[width=2in]{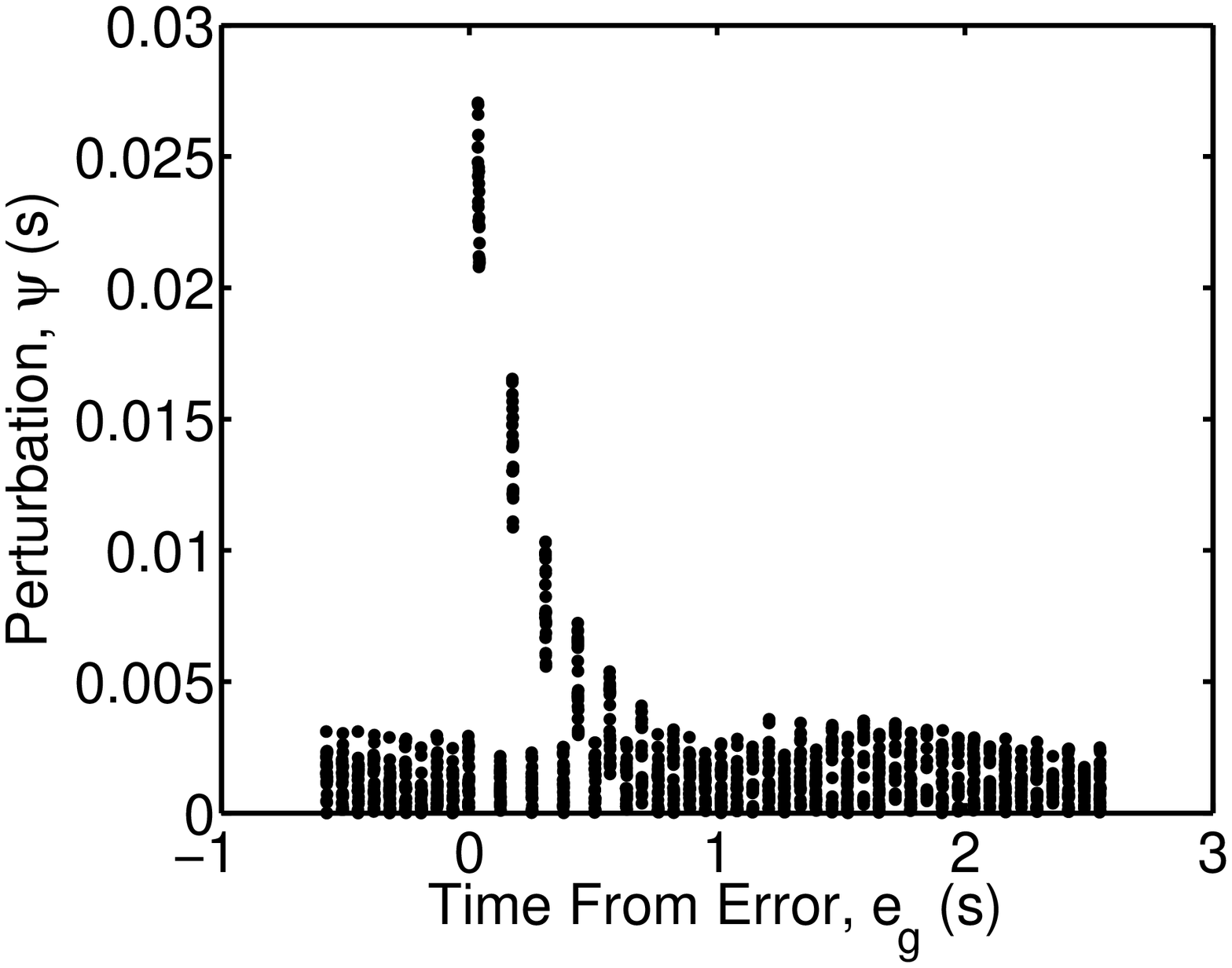} \\
\raisebox{1.35in}{\textsf{\large C}} &
\includegraphics[width=2in]{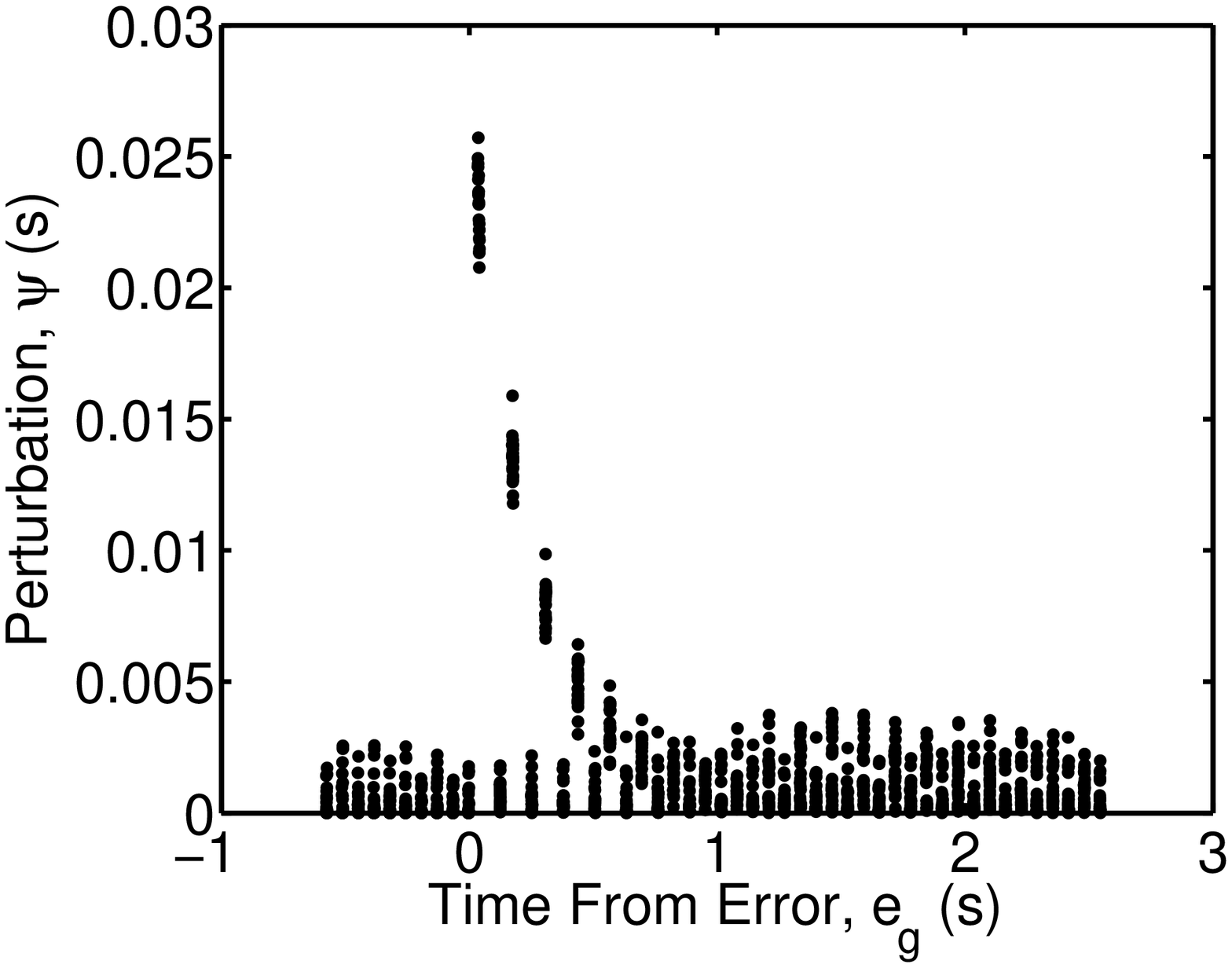} &
\raisebox{1.35in}{\textsf{\large D}} &
\includegraphics[width=2in]{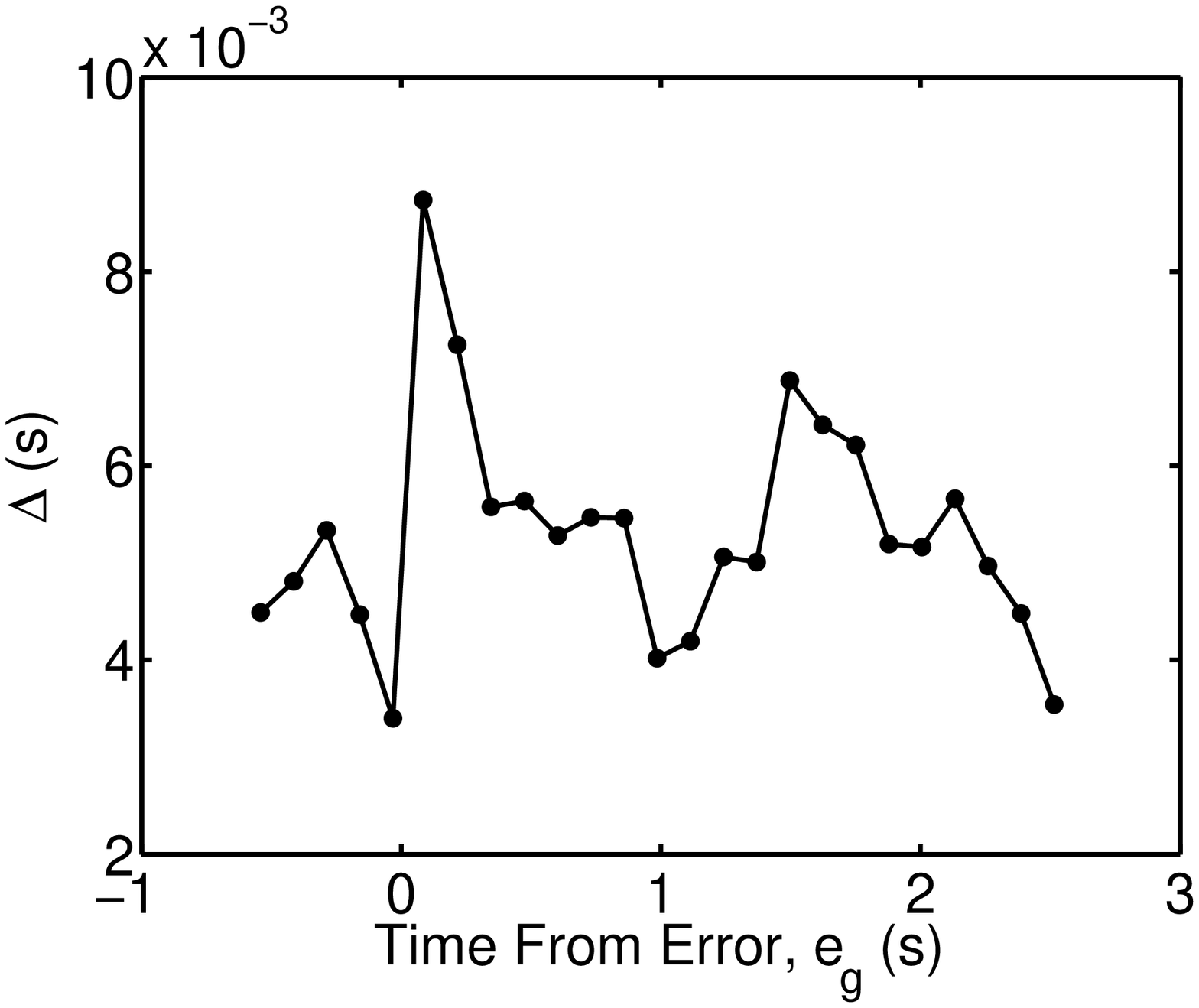}
\end{tabular}
\end{center}
\caption{Error responses for windowed quasiperiodic behavior
($I=63.67$ms, $N/I=1.64$). Phase return map (A) is a 1D curve. High
precision (B) and low precision (C) perturbation plots, Note linear Y
axis for recovery plot (D). ($m=49$ errors, $c=25$
spikes/error)\label{fg:wind-quasi}}
\end{figure}

Figure~\ref{fg:wind-quasi} presents results for a high-input-rate
($N/I>1$) quasiperiodicity. As can be seen in the phase return map
(A), the model only fired within a relatively restricted range of
phases, approximately 59--63ms after a presynaptic spike. This is an
example of a \emph{windowed} behavior~\cite{segundo-etal91a}, in which
the postsynaptic cell is only able to fire within some narrow
time. Both high (B) and low (C) precision had a spike timing
variability of around 3ms before an error, with a peak perturbation of
28ms and perturbation range of about 8ms after an error. The range of
differences (D) was less than doubled by an error, with an oscillatory
recovery with period a bit over 1s and apparent complete recovery
after 4s.

\subsection{Chaos}

\begin{figure}
\begin{center}
\begin{tabular}{lclc}
\raisebox{1.35in}{\textsf{\large A}} &
\includegraphics[width=2in]{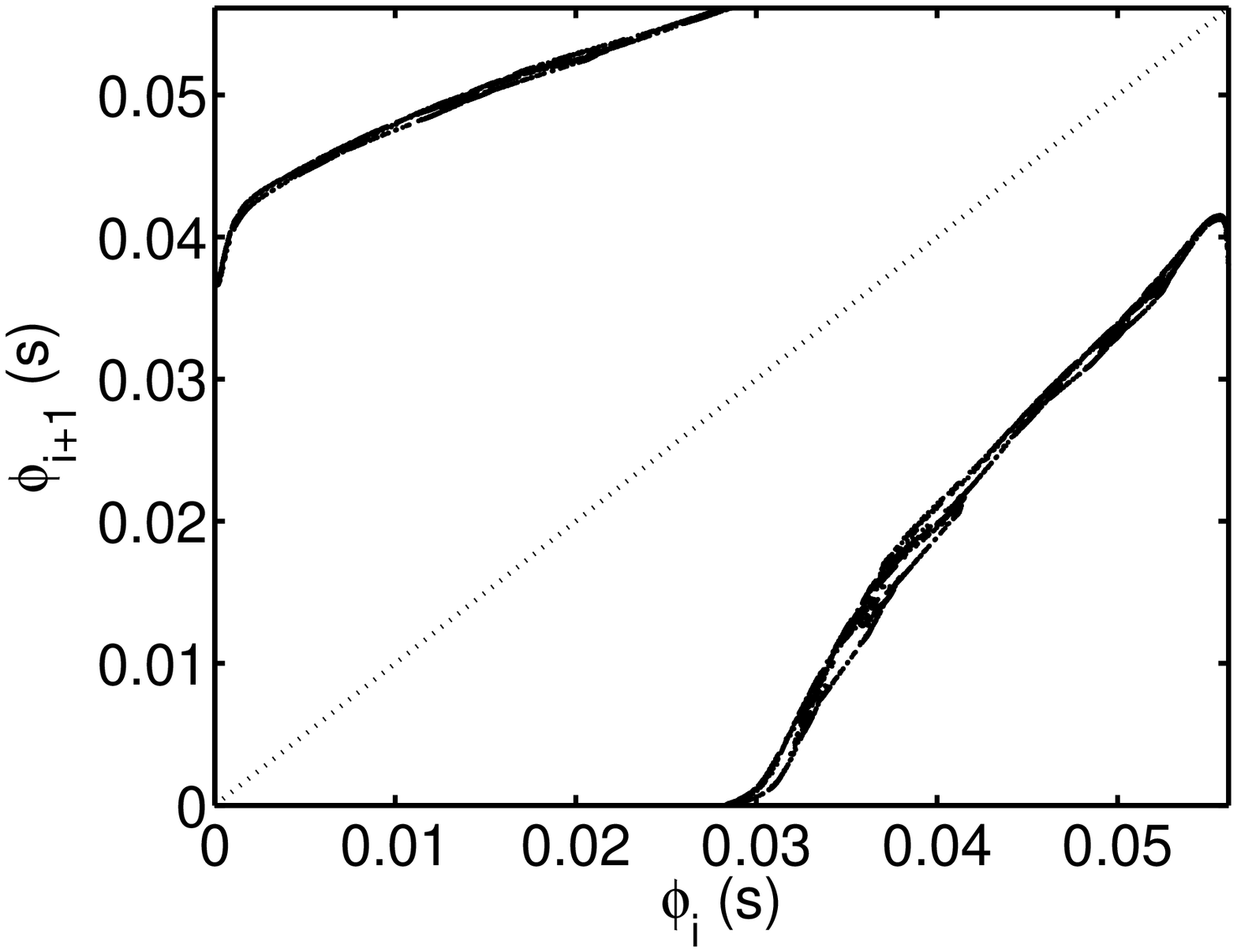} &
\raisebox{1.35in}{\textsf{\large B}} &
\includegraphics[width=2in]{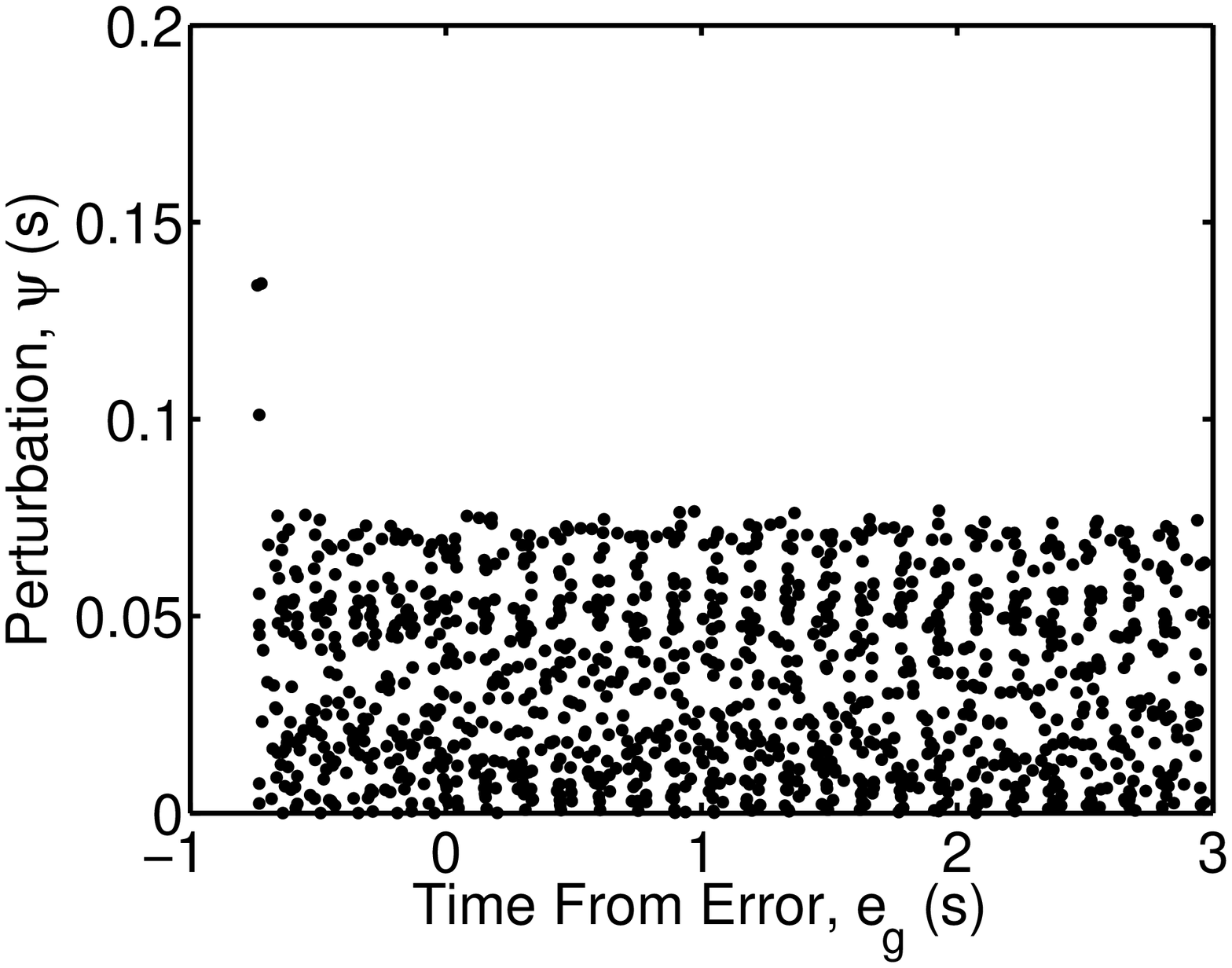} \\
\raisebox{1.35in}{\textsf{\large C}} &
\includegraphics[width=2in]{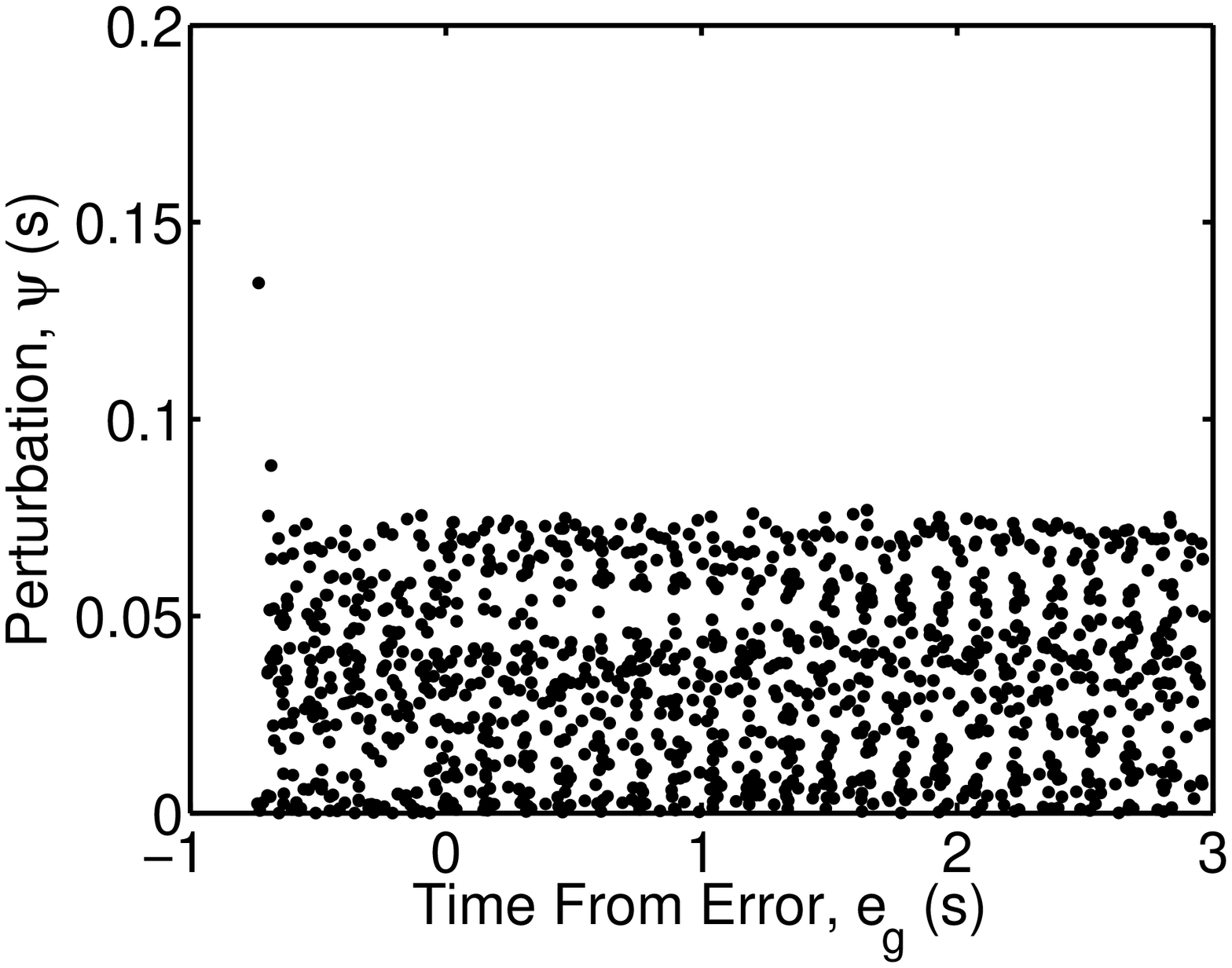} &
\raisebox{1.35in}{\textsf{\large D}} &
\includegraphics[width=2in]{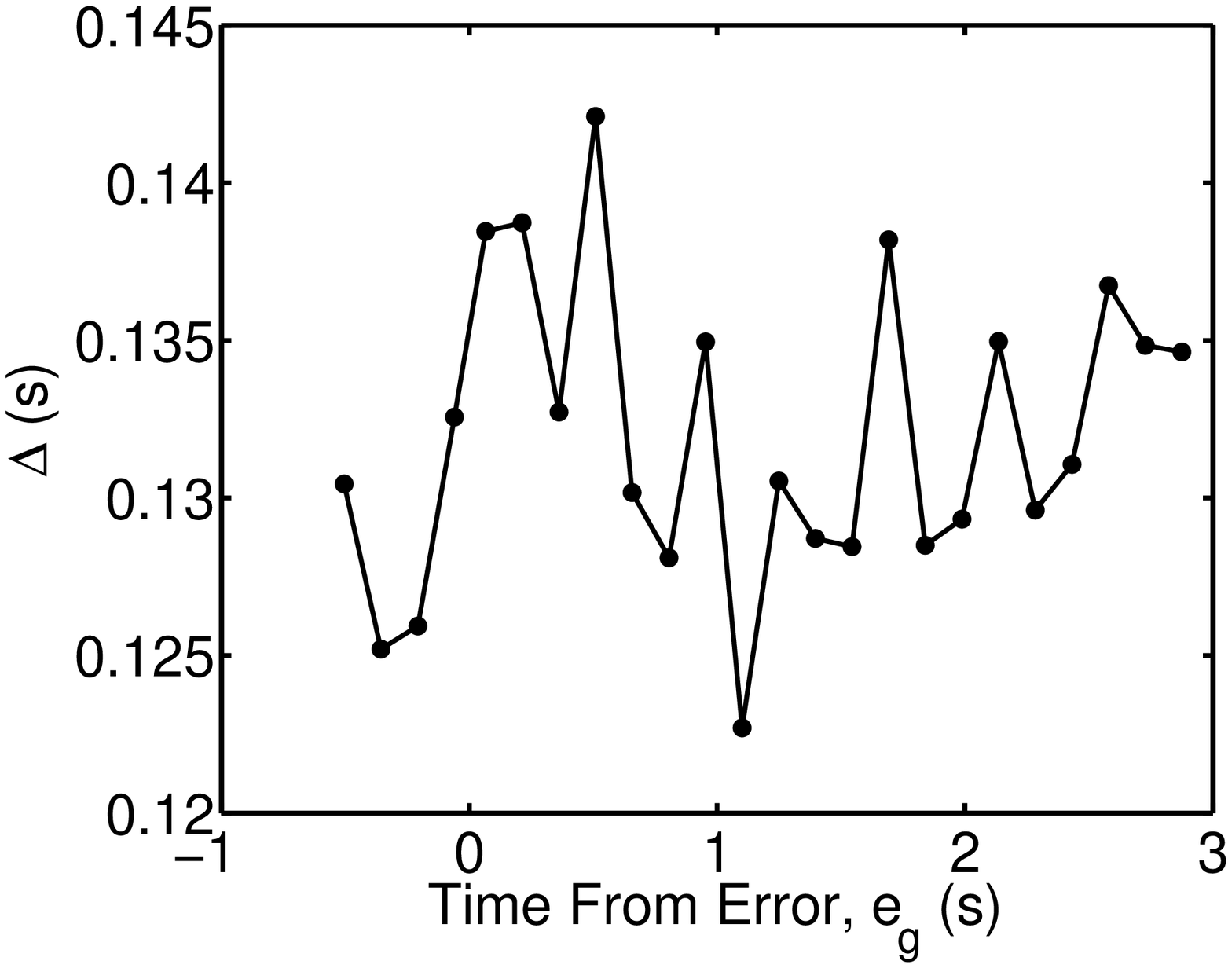}
\end{tabular}
\end{center}
\caption{Chaotic behavior ($I=56.14$ms, $N/I=1.86$). Phase return map
(A) is not a 1D curve. High precision (B) and low precision (C)
perturbation plots. Note linear Y axis for recovery plot
(D). ($m=49$ errors, $c=25$ spikes/error)\label{fg:chaos}}
\end{figure}

Figure~\ref{fg:chaos} shows the error responses for a high-input-rate
chaotic behavior. The phase return map (A) is ``braided;'' clearly not
a 1D curve. The neuron could fire between 0 and 56ms ($I$) after a
presynaptic spike, with a postsynaptic interval range of 139--155ms
and mean of 148ms. So, this response was not windowed. As with the
low-input-rate quasiperiodic behavior, there was little difference
between high (B) and low (C) precision responses --- in both cases,
postsynaptic spike timing variability decreased after the error. There
was perhaps a slight increase in the recovery plot (D) after an error,
with subsequent decrease and increase again, but this was not dramatic.

\subsection{Bifurcation Boundaries}

\begin{figure}
\begin{center}
\begin{tabular}{rl}
\raisebox{1.75in}{\textsf{\large A}} &
\includegraphics[width=2.8in]{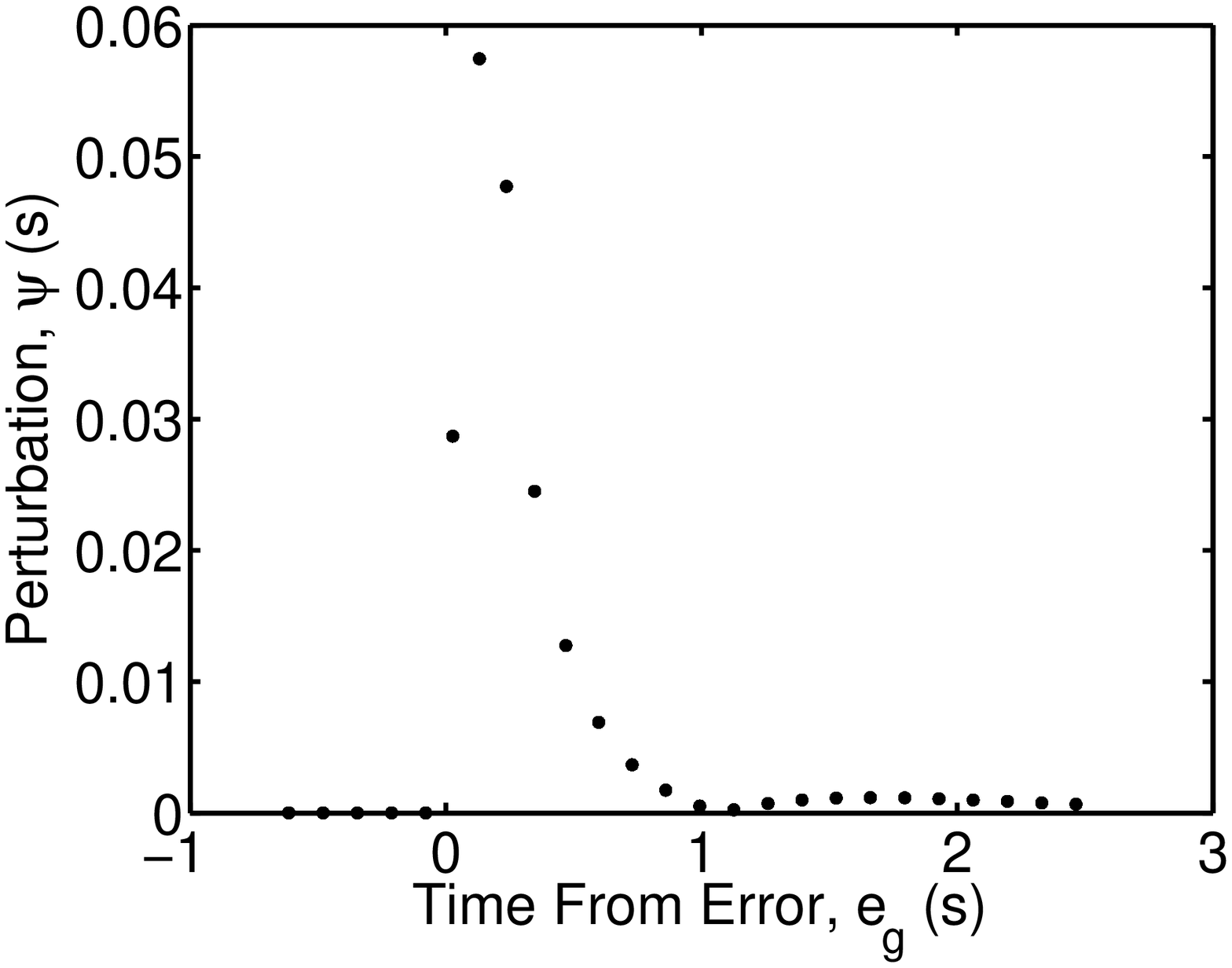} \\
\raisebox{1.75in}{\textsf{\large B}} &
\includegraphics[width=2.8in]{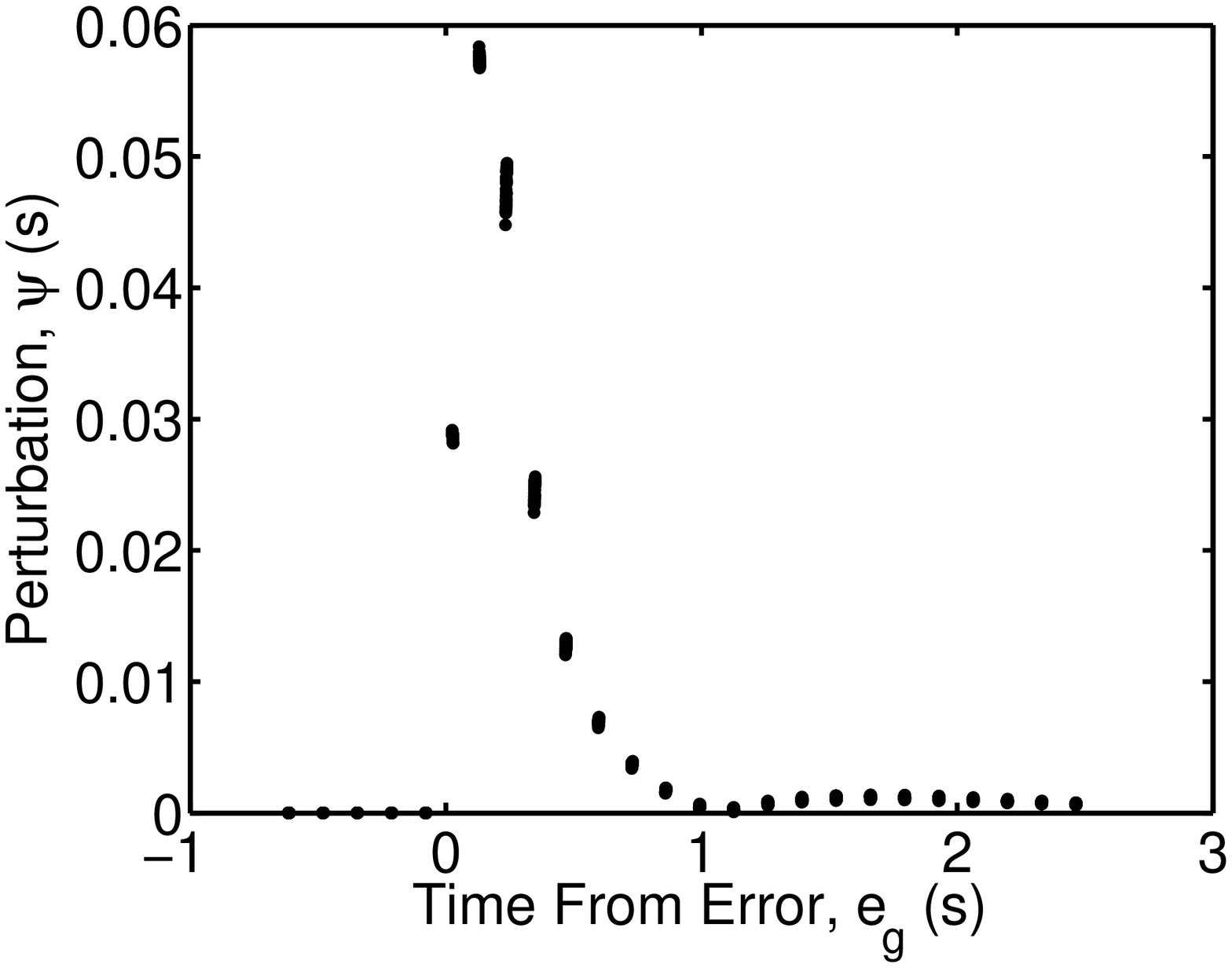} \\
\raisebox{1.75in}{\textsf{\large C}} &
\includegraphics[width=2.8in]{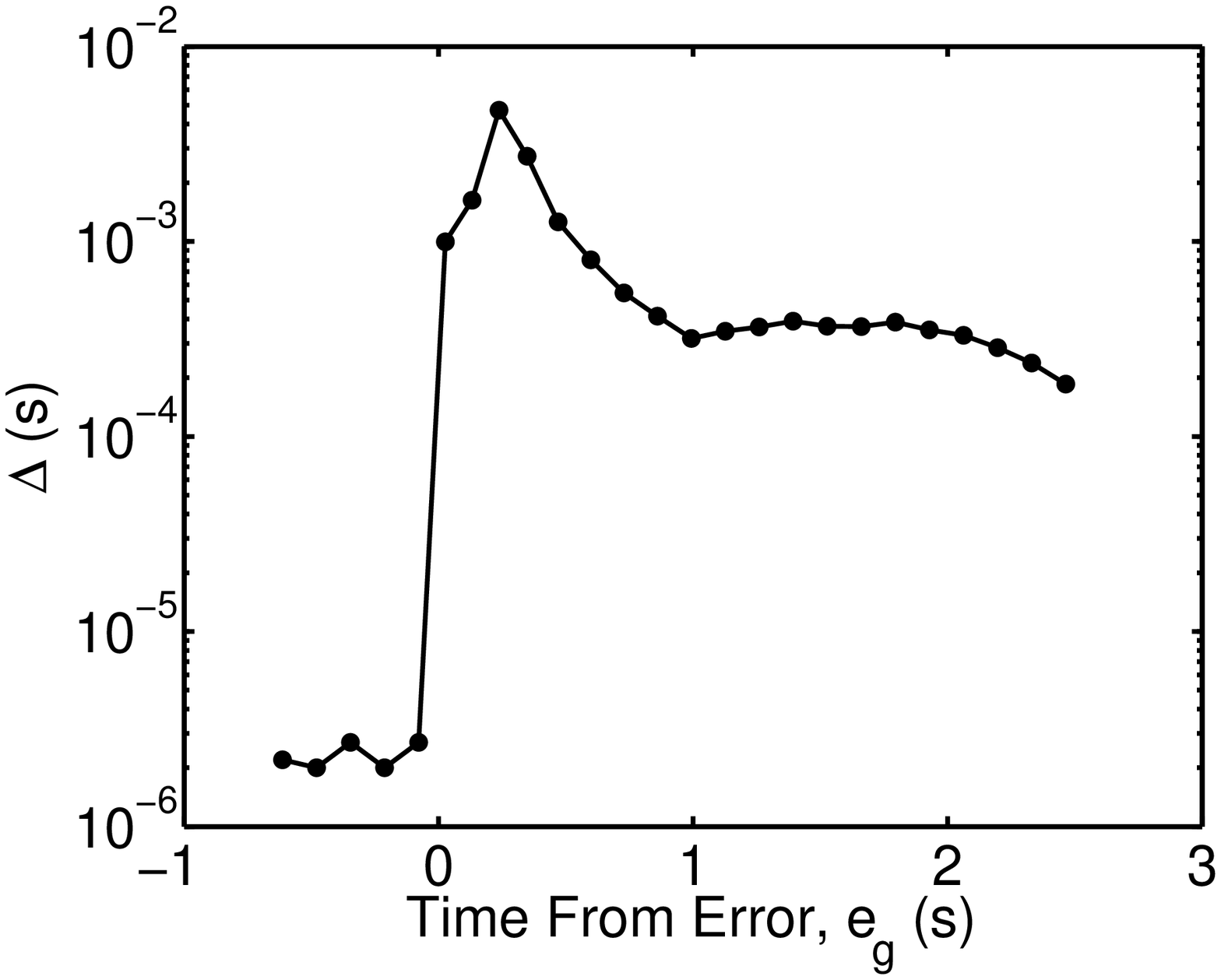}
\end{tabular}
\end{center}
\caption{Results for edge of \ratio{1}{1}
locking ($I=133.88$ms, $N/I=0.78$). High precision perturbation graph
(A) shows identical responses to all errors; low precision (B) shows
range of responses. Both differ from farther away from the bifurcation
(Figure~\protect\ref{fg:1to1}). Recovery plot (C) shows greatly
extended recovery. ($m=49$ errors, $c=25$
spikes/error)\label{fg:1to1edge}}
\end{figure}

The earlier locking examples were chosen to lie within the middle of
the range of presynaptic rates that elicited the behavior. For input
rates near an extreme presynaptic rate, different results were
found. This is illustrated in Figure~\ref{fg:1to1edge} for an input
near the low rate bifurcation point bounding \ratio{1}{1} locking. As
in the middle of the range, high precision inputs (A) elicited the
same response to every error, with a maximum perturbation of almost
60ms. For low precision inputs (B), maximum error perturbations were
between 55ms and 60ms. Unlike the previous \ratio{1}{1} example, the
peak perturbation was reached slowly --- after three postsynaptic
spikes --- and the recovery was more gradual (C), with a broad
secondary peak around $e_g=1.5$s and a value still about two orders of
magnitude greater than that before the error at $e_g=2.5$.

\section{Discussion}

This paper has examined the interaction of neuron dynamics and spike
timing precision in error correction in a model of the living
prototype of an inhibitory synapse and postsynaptic neuron. The
dynamical behaviors included locked, quasiperiodic, and chaotic;
additional issues, such as proximity to bifurcation points and
windowing, were touched upon. All analysis was based on spike times
and interspike intervals, rather than state space analysis. This is
especially significant, in the author's view, because that is exactly
the message that neurons transmit to each other. Thus, variations seen
here are at least potentially significant (as opposed to changes in
internal state that might have no impact on spike timing). 

It is important to remember that, in all cases, high precision
erroneous simulations were compared to high precision error free ones
and low precision erroneous simulations were compared to low precision
error free. Thus, when a recovery plot was generated, values larger
than those just before an error showed that at least some low
precision responses were significantly farther away from ``full
recovery'' than the high precision ones.

Among all behaviors examined, phase lockings produced the most rapid
recovery, with lower ratios that occur in broader rate domains
(\ratio{1}{1} and \ratio{1}{2}) recovering faster. One reason for this
likely was the gradient of the attractor basin surrounding such
behaviors. During the acceptably exponential phase of
their recoveries (roughly from 0.5s to 1s after an error),
\ratio{1}{1} and \ratio{1}{2} locking had similar time
constant (0.20s and 0.27s), despite the fact that their associated
presynaptic rates differed by more than a factor of two. The
conclusion is that, at least for that phase of error recovery, it was
\emph{not} the case that each presynaptic spike simply corrected the
error by some amount proportional to the perturbation. Rather, the
gradients of the two basins of attraction --- at least at those distances
from the attractors --- were similar. Based on computed recovery time
constants, the worst case low precision recoveries had significantly
longer recovery times --- on the order of seconds longer.

For \ratio{2}{3} locking and all nonlocked responses, the ``excess
perturbation'' of low precision versus high precision (the peak of the
recovery plot) was much less than for the lower ratio locked
behaviors. In fact, for \ratio{2}{3} locking, it was roughly a factor
of five greater, while for all nonlocked behaviors it was less than a
factor of two. There were two possible mechanisms underlying this: a
narrow rate domain that elicited the particular behavior and
desynchronization of erroneous and error free responses by errors.

The first matter, that of domain width, was especially relevant to
\ratio{2}{3} locking. As can be seen by comparing
Figure~\ref{fg:2to3}(C) to Figures~\ref{fg:1to1}(C)
and~\ref{fg:1to2}(C), the peak was not too different than the lower
ratio lockings; the difference was the degree of perturbation before
the error (more than two orders of magnitude higher in the case of
\ratio{2}{3} locking, roughly at the level of the added low precision
jitter). This was probably caused by the jitter in the low precision
simulations bumping the state out of the locking behavior
\cite{beierholm-etal01,tiesinga-sejnowski01}, causing the two low
precision simulations to remain partially desynchronized --- they
were, at least part of the time, away from the synchronizing effect of
the periodic attractor (see discussion below of bifurcation boundary
transitions).

That was a ``minor'' desynchronization effect, because there was a
periodic attractor which tended to pull the erroneous and error free
simulations' states back together. Such was not the case for aperiodic
behaviors. In those cases (quasiperiodicity and chaos), once the
neuron had been perturbed by the first error, returning to the
attractor did not bring its subsequent spikes back to the same times
they would have occurred in the absence of the error. This was the
case for both high and low precision. For the quasiperiodic cases
(Figures~\ref{fg:quasiperiodic} and~\ref{fg:wind-quasi}), there were
``preferred'' firing intervals (where the phase return map was closest
to the diagonal in both figures, and additionally due to windowing
itself for windowed quasiperiodicity). The closer the quasiperiodicity
came to being periodic (the more dominant the preferred firing
intervals were), the more it would act to resynchronize the erroneous
simulation with the error free one.

For Figure~\ref{fg:quasiperiodic}, this was not very close at all
(about 25ms), the erroneous and error free experiments were completely
desynchronized by the first error, and as a result their ``perturbations''
in Figures~\ref{fg:quasiperiodic}(B) and~(C) before the error could be
as large as the full range of possible phase differences
($I/2$). Additionally, the high and low precision cases thereafter
were also decorrelated, and so the range of their perturbation differences
in~(D) was the sum of their maxima. However, in both the high and low
perturbation cases, errors arrived at approximately the same time
(within $\pm wI$ of each other), and so the errors paradoxically
served as a partial resynchronizing input, as indicated by the decrease
in~(D) after the error.

Windowed quasiperiodicity, however, behaved more like a weak
locking. As Figure~\ref{fg:wind-quasi} shows, the neuron was
constrained to fire within a 4ms window relative to an input spike
when stationary~(A), which was much less than the 25--30ms
perturbation induced by an error~(B, C). As a result, the cell
exhibited an oscillatory recovery, with high precision input being
somewhat more effective~(D).

Chaos, by definition of its hallmark sensitive dependence on initial
conditions, was the extreme case of these desynchronizing processes,
with the jitter of low precision leading those cases to diverge
rapidly from high precision and the presence of errors having a
similar effect between erroneous and error free experiments. As
Figure~\ref{fg:chaos}(D) shows, error aftereffects were qualitatively
similar to (though perhaps more extreme than) that for
quasiperiodicity in Figure~\ref{fg:quasiperiodic}(D), with the
similarly-timed errors affecting the high and low precision erroneous
experiments in similar fashion.

The final results in Figure~\ref{fg:1to1edge} address the issue of
proximity to a bifurcation boundary. This is also relevant to the
effects of domain width and similarity or dissimilarity of attractor
basins and attractors on error recovery. In both the high and low
precision cases, in the absence of errors the neuron's state was on or
near a period-one attractor at all times (the jitter was not large
enough to destroy locking, as was verified by examination of the phase
return map for the low precision case). Unlike the simulation in
Figure~\ref{fg:1to1}, however, the perturbation produced by an error
did \emph{not} displace the state into the basin of another
\ratio{1}{1} locking. Instead, the perturbation moved the state across
the bifurcation boundary into the basins of qualitatively different
attractors, and its motion thereafter was very different than that
of recovery exclusively through period-one attractor basins.

This raises an important point regarding the nature of the
perturbation produced by an error. Because the presynaptic intervals
served as a control parameter, errors were not merely perturbations of
the neuron's state within its state space, $\Sigma \subset
\IR^n$. Instead, they were perturbations within the system's
\emph{response space}, $I \times
\Sigma$ \cite{thom75,abraham-shaw84}. The perturbation of an ongoing
behavior near a bifurcation boundary or within a narrow domain (such
as in Figures~\ref{fg:2to3}--\ref{fg:1to1edge}) displaced the neuron's
state across domains of different behaviors.  After the perturbation,
the re-established sequence of spike times drew the neuron through its
response space and back across (possibly multiple) bifurcations into
basins of attractors qualitatively like its ``home'' (stationary
attractor). Its evolution, however, because of these dissimilar
basins, didn't necessarily lead it toward the region of $\Sigma$ that
contained this home. As a result, the perturbation and recovery plots
could show multimodal, or oscillatory, shapes.

Having established that even a small amount of spike timing jitter can
have observable effects on the time course of error recovery (and
potentially on the activity of further postsynaptic neurons),
subsequent work is progressing along a number of fronts:
\begin{itemize}
\item A more complete model of the bifurcation behavior of
error recovery \cite{stiber-holderman04a} is being developed. Rather
than using some property of the stationary behavior of a dynamical
system, one can examine how properties of transient neural responses
depend on pre-existing stationary behaviors.

\item The role of degree of spike timing accuracy on error is being
investigated \cite{stiber-holderman04b}. The current results show that
even small amounts of jitter can have a significant effect on error
recovery.  The amounts used here were in the range of 1--4ms or so,
and thus are relevant to a range of biological systems, based on
observations (for multiple spike trains) of correlations and
correlation-dependent activity in retinal ganglion cells and lateral
geniculate cells, and are at the low end for some cortical cells
\cite{bair99}.  Preliminarily, order of magnitude jitter increases in
general produce similar effects, though, as might be expected in a
nonlinear system, linear increases do not appear to produce linear
changes in recovery.

\item The transient response of this model is being examined in terms
of its internal state variables \cite{stiber-pottorf04}, as a
preliminary to developing a simplified model that captures the error
responses and is more amenable to analytic approaches. Early results
suggest that spike rate adaptation is the most significant contributor
to the observed error responses.
\end{itemize}

A final matter is that of the application of Shannon's information
theory to coding by nonlinear dynamical neurons. Techniques from the
field of information theory
\cite{shannon48,cover-thomas91} are currently of great interest in
addressing the question of what spike trains can be said to carry
identical information
\cite{bialek-etal91,borst-theunissen99,fuhrmann-etal02}, and thus
whether the timing of individual spikes or average firing rate is
relevant.  Entropy calculations can be quantified in terms of bits of
information per spike --- a compelling statistic. The typical
experimental approach is to present a stimulus multiple times and use
the ``average'' response (where this averaging might be accomplished
in varying ways) to compute entropy or mutual information. This is
then used to infer characteristics of the neural code (e.g., rate
versus time codes).

It is important, however, to remember that information theory tells
one nothing about the semantics of a code
\cite{shannon-weaver49}. Primarily, and in conjunction with knowledge
of the noise in a communication channel, it addresses issues of
information capacity: how close to optimally compressed a signal
is. This part of Shannon's work --- the Noisy Channel Coding Theorem
--- is usually used in neuroscience only in terms of estimating
``channel noise'' from multiple presentations of the same
stimulus. However, another application of this theorem is the creation
of codes that use more bits (symbols, or in the context of neural
coding, greater spike timing precision) than the required minimum to
provide for error correction in the presence of channel noise.

As these results clearly show, both the quantitative and qualitative
results of a presynaptic error depend critically on a neuron's
internal state's ongoing evolution. It is a mapping $f: I \times
\Sigma \rightarrow O \times \Sigma$, from input and state to
output and new state. This suggests that the computation performed by
a neuron will inevitably result in a decrease in mutual information
between its input and output, according to the Data Processing Theorem
\cite{shannon48,cover-thomas91}. In contrast, Shannon's formal
definition of an encoder or decoder is a stationary mapping $f: A^*
\rightarrow B^*$ from input sequences of symbols taken from some set
$A$ to output sequences of symbols from some (possibly different) set
$B$. In other words, Shannon's information theory applies when the
encoding of an input symbol depends only on its context within the
sequence of input symbols. On the other hand, for a nonlinear
dynamical system, output depends on input \emph{and state}, where
state is not merely a function of input but also of intrinsic
dynamics.  Therefore,
\emph{Shannon's information theory is conceptually incompatible with
neural information processing}. In practical terms, this implies that
information theoretic techniques can only be used to establish lower
bounds on matters such as spike timing precision.

\appendix
\section{SAO Model Equations}
\label{sc:eqns}

The SAO model equations used here are documented more completely in
\cite{edman-etal87,stiber-etal97b}; the following is a brief
summary. It is a permeability-based, rather than conductance-based,
model with the following main equations:

\begin{eqnarray}
\deriv{V_{m}}{t} & = & -(\ix{Na} + \ix{K} + \ix{L,Na} + \ix{L,K} +
\ix{L,Cl}
   +\ix{p} + \ix{bias} + \ix{syn})/C_{m}          \label{eq:pv} \\
\ix{Na}  & = & A \pbx{Na} m^{2} h l \frac{V_{m}F^{2}}{RT}
   \frac{[\Na]_{o} - [\Na]_{i} \exp{FV_{m}/RT}}{1 - \exp{FV_{m}/RT}}
                                                  \label{eq:ina} \\
\ix{K}   & = & A \pbx{K} n^{2} r \frac{V_{m}F^{2}}{RT}
   \frac{[\K]_{o} - [\K]_{i} \exp{FV_{m}/RT}}{1 - \exp{FV_{m}/RT}}
                                                  \label{eq:ik} \\
\ix{L,X} & = & A \px{L,X} \frac{V_{m}F^{2}}{RT}
   \frac{[X]_{o} - [X]_{i} \exp(FV_{m}/RT)}{1 - \exp(FV_{m}/RT)}
                                                  \label{eq:il} \\
\ix{p}   & = & \frac{AF}{3}
   \frac{\jbx{p,Na}}{\left( 1 + \frac{K_{m}}{[\Na]_{i}} \right)^{3}}
                                                  \label{eq:ip} \\
\ix{syn} & = & A \px{syn} \frac{V_{m}F^{2}}{RT}
   \frac{[\Cl]_o - [\Cl]_i \exp(-FV_m/RT)}{1 - \exp(-FV_m/RT)}\\ 
\label{eq:isyn}
\nonumber
 & & \times\sum_{\forall s_k \leq t} \left(e^{(s_k-t)/\tau_+} -
      e^{(s_k-t)/\tau_-}\right)
\end{eqnarray}

\begin{longtable}{|l|r|l|}
\caption{Constants used in this simulation, in rough order of
appearance in equations in this appendix.\label{tb:constants}} \\ \hline
\bf Constant & \bf Value & \bf Units (MKS) \\ \hline
\endfirsthead
\caption[]{Constants used in this simulation, cont'd. Note that the
$V_m$ here is the voltage associated with the $m$ gating variable, not
the membrane potential.} \\ \hline
\bf Constant & \bf Value & \bf Units (MKS) \\ \hline
\endhead
\hline
\endfoot
$C_m$     & $7.5 \times 10^{-9}$   & Farads          \\
\ix{bias} & $-2.5 \times 10^{-9}$  & A   \\
$A$       & $1.0 \times 10^{-7}$   & \spscript{m}{2} \\
\pbx{Na}  & $6.0 \times 10^{-6}$   & m/s \\
$[\Na]_{o}$ & 325.0                & mM  \\
$[\Na]_{i}$ & 10.0                 & mM  \\
$T$       & 291.15                 & $^\circ$K \\
\pbx{K}   & $2.0 \times 10^{-6}$   & m/s \\
$[\K]_{o}$  & 5.0                  & mM  \\
\px{L,Na} & $5.0 \times 10^{-10}$   & m/s \\
\px{L,K}  & $1.8 \times 10^{-8}$   & m/s \\
\px{L,Cl} &  $1.1 \times 10^{-9}$  & m/s \\
$[\Cl]_{o}$ & 650.0                & mM  \\
$[\Cl]_{i}$ & 46.0                 & mM \\
\jbx{p,Na}& $6.0 \times 10^{-6}$   & mol/(\spscript{m}{2}s) \\
$K_{m}$   & 13.4                   & mM  \\
\px{syn}  & $6.0 \times 10^{-8}$   & m/s \\
$\tau_+$  & $0.25 \times 10^{-3}$  & s   \\
$\tau_-$  & $0.5 \times 10^{-3}$  & s   \\
$\nu_m$      & 0.0                 & dimensionless \\
$\nu_h$      & 0.0                 & dimensionless \\
$\nu_l$      & 0.0                 & dimensionless \\
$\nu_n$      & 0.03                & dimensionless \\
$\nu_r$      & 0.5                 & dimensionless \\
$z_m$       & -3.1                 & dimensionless \\
$z_h$       & 4.0                  & dimensionless \\
$z_l$       & 3.5                  & dimensionless \\
$z_n$       & -2.6                 & dimensionless \\
$z_r$       & 4.0                  & dimensionless \\
$V_m$        &    $-19.0 \times 10^{-3}$   &   V \\
$V_h$        &    $-35.0 \times 10^{-3}$   &   V \\
$V_l$        &    $-53.0 \times 10^{-3}$   &   V \\
$V_n$        &    $-18.0 \times 10^{-3}$   &   V \\
$V_r$        &    $-56.0 \times 10^{-3}$   &   V \\
$\overline{\tau}_m$  & $0.3 \times 10^{-3}$     & s \\
$\overline{\tau}_h$  & $5.0 \times 10^{-3}$     & s \\
$\overline{\tau}_l$  & $1700.0 \times 10^{-3}$  & s \\
$\overline{\tau}_n$  & $6.0 \times 10^{-3}$     & s \\
$\overline{\tau}_r$  & $2000.0 \times 10^{-3}$  & s \\
\end{longtable}

Values for constants are given in Table~\ref{tb:constants}. Besides
active and leak \Na{}, \K{}, and \Cl{} currents, there is a bias
current (\ix{bias}), generated in response to muscle stretch, an
active pump (\ix{p}), and a synaptic current (\ix{syn}). There are
five gating variables, $m$, $h$, $l$, $n$, and $r$, with first-order
kinetics described by~(\ref{eq:g}--\ref{eq:qg}) (where $g \in \{m, h,
l, n, r\}$).

\begin{eqnarray}
\deriv{g}{t} & = & (g_{\infty} - g) \frac{1}{\tau_g} \label{eq:g} \\
g_{\infty} & = & \nu_{g} + \frac{1 - \nu_{g}}{1 +
\exp\left[\frac{z_{g} e}{kT} (V_{m} - V_{g})\right]} \label{eq:gi} \\
\tau_{g} & = & \frac{Q_{g} \bar{\tau}_{g}}{\exp\left[\frac{\delta_{g}
z_{g} e}{kT} (V_{m} - V_{g})\right] + \exp\left[\frac{(\delta_{g}-1)
z_{g} e}{kT} (V_{m} - V_{g})\right]} \label{eq:tg} \\
Q_{g} & = & \left(\frac{1 -
\delta_{g}}{\delta_{g}}\right)^{\delta_{g}} + \left(\frac{1 -
\delta_{g}}{\delta_{g}}\right)^{\delta_{g}-1} \label{eq:qg}
\end{eqnarray}

\bibliographystyle{apalike}
\bibliography{short-strings,stiber,neuro,scientific-computing,dynamics,signal-processing}

\end{document}